\RequirePackage{ifpdf}
\documentclass[11pt,letterpaper]{JHEP3}
\usepackage{epsfig}
\usepackage[latin1]{inputenc}
\usepackage{bbold,amsfonts}
\usepackage{graphicx}
\usepackage{amssymb,amsmath}
\usepackage{fancybox,framed}
\usepackage{dsfont}
\usepackage{mathtools}
\usepackage{braket}
\usepackage{slashed}
\usepackage[makeroom]{cancel}

\author{Lorenzo~Bianchi$^{a}$, Marco S. Bianchi$^{b}$\\\\
$^{a}$ Institut f\"ur Physik,
Humboldt-Universit\"at
zu Berlin\\ Zum Gro\ss en Windkanal 6, 12489 Berlin, Germany
\\
$^{b}$
Centre for Research in String Theory,
School of Physics and Astronomy\\
Queen Mary University of London,
Mile End Road, London E1 4NS, UK \\
\qquad\\
E-mail: \email{lorenzo.bianchi@physik.hu-berlin.de, m.s.bianchi@qmul.ac.uk}
}

\abstract{We initiate the perturbative study of the S-matrix for the excitations on top of the GKP vacuum at strong coupling. Using the string sigma model action expanded around the null cusp classical solution, we compute the tree-level S-matrix elements and compare them with the predictions from the asymptotic Bethe Ansatz. We also check the factorization of the three-body S-matrix for various bosonic processes, finding precise agreement with the constraints imposed by integrability.
}

\preprint{September 2015\\HU-EP-15-40/\\QMUL-PH-15-16
}

\title{Worldsheet scattering for the GKP string}


\csname @addtoreset\endcsname{equation}{section}

\def\bseq{\begin{subequation}}  
\def\eseq{\end{subequation}}
\def\bsea{\begin{subeqnarray}}  
\def\esea{\end{subeqnarray}}

\newcommand{\beq}{\begin{equation}}
\newcommand{\bea}{\begin{eqnarray}}
\newcommand{\eea}{\end{eqnarray}}
\newcommand{\eeq}{\end{equation}}

\newcommand{\g}{\gamma}

\newcommand{\D}{\Delta}

\newcommand{\p}{\pi}
\renewcommand{\P}{\Pi}

\newcommand{\ppp}{\text{p}}

\def\beq{\begin{equation}}
\def\eeq{\end{equation}}
\def\bea{\begin{eqnarray}}
\def\eea{\end{eqnarray}}

\def\g{\gamma}

\def\th{\theta}

\def\D{\Delta}

\declareslashed{}{\backslash}{0}{0}{\Omega}
\declareslashed{}{\backslash}{0}{0}{O}
\begin{document}

\allowdisplaybreaks

\section{Introduction}

More and more aspects of four-dimensional ${\cal N}=4$ SYM in the planar limit have been revealed to be deeply connected to physics in two dimensions.
On the one hand the AdS/CFT correspondence \cite{Maldacena:1997re} relates its strong coupling limit to a superstring theory defined on a two-dimensional worldsheet.
On the other hand an increasing number of quantities of ${\cal N}=4$ SYM have been shown to be computable at any coupling via a description in terms of an integrable spin chain \cite{Minahan:2002ve,Beisert:2003yb}.

A corner of this big picture which we will focus on in this paper is the integrability of the twist-two operators of planar ${\cal N}=4$ SYM.
They belong to the $sl(2)$ sector of the single trace operators of the theory and in the large spin limit their anomalous dimensions are fully determined by a set of asymptotic Bethe Ansatz (ABA) equations \cite{Beisert:2005fw,Beisert:2006ez}.
In the strong coupling regime, via the AdS/CFT correspondence, such twist-two operators are conjectured to be dual to a folded string spinning around its center of mass in $AdS_3 \subset AdS_5$ \cite{Gubser:2002tv,Frolov:2002av}, whose large spin limit is a fairly simple string solution, amenable of detailed analyses.

One equivalent and fruitful way of describing this system is the light-cone gauge-fixed Lagrangian of the $AdS_5\times S^5$ string sigma model \cite{Metsaev:2000yu,Metsaev:2000yf} expanded around the null cusp background \cite{Giombi:2009gd}. 
From the quadratic part of the Lagriangian it is possible to read out the spectrum of the excitations of the model at infinite coupling.
These are a mass $\sqrt{2}$ complex scalar $x$, a mass 2 scalar $\phi$, $8$ mass 1 fermions and five massless scalars.
This spectrum is in partial agreement with the degrees of freedom of the Bethe equations valid at any coupling, with discrepancies connected to the nonperturbative dynamics of the $O(6)$ sigma model emerging at strong coupling in the Alday-Maldacena limit \cite{Alday:2007mf}.
Nevertheless, the light-cone gauge-fixed Lagrangian can be taken as the starting point for performing perturbation theory and computing quantities of interest at strong coupling. In particular, the light-cone gauge choice makes the Feynman rules fairly simple, so that this Lagrangian is suitable for computing quantum corrections.
This approach has been applied to the study of the free energy of the theory \cite{Giombi:2009gd} which is dual to the anomalous dimension of a cusped light-like Wilson line in planar ${\cal N}=4$ SYM at strong coupling. Such a computation has been pushed to two-loop order and agrees with the ABA prediction \cite{Basso:2007wd} providing one of the most spectacular mutual tests of integrability at strong coupling and of the AdS/CFT correspondence (see also \cite{Roiban:2007jf,Roiban:2007dq,Roiban:2007ju}).

The ABA also allows to compute the momentum of the excitations of the GKP string, and consequently determine their dispersion relations to all orders.
Again, these predictions from integrability were compared at next-to-leading order at strong coupling by computing the two-point functions of the worldsheet excitations.
Interestingly, as mentioned above, the agreement in this case is only partial and the reasons for the mismatches were clarified in \cite{Zarembo:2011ag}. In particular it was shown that perturbation theory within this model can fail to produce sensible results for particular quantities, due to the onset of nonperturbative effects.

We flash that a similar ABA description also exists for the $AdS_4\times\mathbb{CP}^3$ GKP string \cite{Gromov:2008qe,Basso:2013pxa}, dual to the large spin limit of twist-one operators of the ABJM superconformal model in three dimensions \cite{Aharony:2008ug}. In this context, starting from the $AdS_4\times\mathbb{CP}^3$ light-cone gauge-fixed Lagrangian \cite{Uvarov:2009hf,Uvarov_main,Uvarov:2011zz}, a parallel computation of the cusp anomalous dimension at two-loops \cite{Bianchi:2014ada}, and of the two-point functions at one loop \cite{Bianchi:2015laa} has been carried out. \\

Integrability is able to provide further fundamental data for solving the GKP string, namely the exact S-matrix for its excitations \cite{Basso:2013pxa,Fioravanti:2013eia,Fioravanti:2015dma}.
This object is interesting per se, since it encloses the dynamics of the model, and additional relevance comes from its remarkable relation to scattering amplitudes of planar ${\cal N}=4$ SYM.
The starting point for building this bridge is a light-like Wilson loop in a conformal gauge theory. One can perform an OPE decomposition of it by selecting two light-like edges, cutting the Wilson loop across them into a bottom and a top part and inserting a basis of eigenstates in the cut \cite{Alday:2010ku,Gaiotto:2011dt}. 
The latter are interpreted as the excitations of the color flux-tube stretching between two null lines. The OPE expansion is then taken by sending to infinity the flux-tube time conjugate to the energy of the excitations. In space-time, this corresponds to flattening the bottom side of the loop, which is in turn equivalent to a multicollinear limit in dual kinematics \cite{Alday:2010ku,Gaiotto:2011dt}.
A generic polygon is fully reconstructed from the OPE decomposition by repeatedly performing the procedure sketched above. This is achieved by dividing the polygon into elementary squares and considering how excitations propagate between two adjacent squares, forming a pentagon, from the bottom to the top edge \cite{Basso:2013vsa}.
The central object enclosing the dynamics of this process has been dubbed the pentagon transition.
The remarkable feature of planar ${\cal N}=4$ SYM is two-fold. On the one hand in this theory the flux-tube excitations are the same as those of the GKP string and their dynamics is completely determined at any coupling by integrability. In particular, the pentagon transitions emerge as ratios of GKP string S-matrix elements.
On the other hand null Wilson loops in ${\cal N}=4$ SYM are dual to scattering amplitudes \cite{Alday:2007hr,Drummond:2007au,Drummond:2007cf,Brandhuber:2007yx}, offering the unprecedented possibility of evaluating the S-matrix of an interacting four dimensional theory at any coupling.
In fact this approach has been applied to and tested against a variety of scattering processes, both at weak and strong coupling \cite{Basso:2013aha,Basso:2014koa,Basso:2014jfa,
Basso:2014nra,Basso:2014hfa,Basso:2015rta,Basso:2015uxa}.
In conclusion there exists a tight interplay between the (flux-tube) S-matrix of the GKP string two-dimensional model and that of the four-dimensional planar ${\cal N}=4$ SYM.\\ 

Recently, the S-matrix for the GKP string has been thoroughly studied using the ABA in \cite{Fioravanti:2013eia,Fioravanti:2015dma}.
This allows to write expressions for its elements, valid at any order.
In particular their expansion at strong coupling in the perturbative regime has been spelled out, which is amenable of perturbative checks.
The aim of this paper is to perform such tests by comparing these integrability based results to the amplitudes which can be computed perturbatively from the light-cone gauge $AdS_5\times S^5$ sigma model lagrangian.
We start by reviewing the action expanded around the cusp background and its Feynman rules.
Next we detail the computation of several S-matrix elements between the particles of the model.
We find that, as long as massless modes do not enter the computation, the results are trustworthy and exhibit perfect agreement with the integrability predictions. In order to make this manifest we express both results in terms of hyperbolic rapidities to allow for comparison.

All other amplitudes, namely that for two fermions and all those with massless scalars as external states, turn out to be troublesome, as might have been expected from the findings of \cite{Giombi:2010bj,Zarembo:2011ag}.
In section \ref{sec:scalars} we comment more extensively on fermion-fermion scattering and propose a trick (though biased by rather strong assumptions) to compute the scalar factor evading the problematic part of the perturbative computation. 

In the last section, as a further check of integrability, we analyse some processes involving six particles in the massive scalar sector of the excitations, namely the scattering of gluons of same helicity and mesons.
These are four possible processes and we verify in all cases that there is no particle production and the S-matrix factorizes. Here we anticipate that the cancellation of the various diagrams in a generic kinematic configuration is considerably more intricate and stunning than the BMN case \cite{Klose:2007rz} due to the presence of cubic and quintic interactions.

\section{The light-cone gauge action}

We work with the light-cone gauge euclidean action of the $AdS_5 \times S^5$ sigma model expanded around the cusp background of \cite{Giombi:2010bj}. We use the version with fermions cast into the Dirac form as in \cite{Zarembo:2011ag}
\begin{equation}\label{eq:action}
S = \frac{T}{2}\int dt \int^\infty_{-\infty} ds\ {\cal L}\qquad\qquad T\equiv \frac{\sqrt{\lambda}}{2\pi}
\end{equation}
where $T$ is the string tension in terms of the ${\cal N}=4$ 't Hooft coupling $\lambda$ and
\begin{align}\label{eq:lagrangian}
{\cal L}  &=
\big|\partial_t x + x \big|^2 +
\frac{1}{z^4} \big| \partial_s x - x \big|^2 + \Big( \partial_t z^M + z^M +
\frac{i}{z^2} \psi^{\dagger}_i \P_{+} (\rho^{MN}){}^i{}_j \psi^j  z_N \Big)^2
+ \nonumber\\
& + \frac{1}{z^{4}} \Big(\partial_s z^M - z^M \Big)^2 + 2\, i\, \psi^{\dagger}_i \partial_t \psi^i - \frac{1}{z^{2}} \Big(\psi^{\dagger}_i \P_{+} \psi^i\Big)^2 + \nonumber\\& + \frac{2i}{z^3}\, \Bigl[-\bar\psi_i \P_{+} (\rho^{\dagger}_6)^{ik} (\rho^M)_{kj} z^M \D_s \psi^j
- \frac{i}{z} (\psi^i)^T \P_{+} (\rho^M)_{ij} z^M \psi^j \D_s x + \nonumber\\& ~~~~~~~~
+ \psi^{\dagger}_i \P_{+} (\rho^\dagger_M)^{ik} z^M (\rho^6)_{kj} \D_s \psi^j
+ \frac{i}{z} \psi^{\dagger}_i \P_{+} (\rho^{\dagger}_M)^{ij} z^M (\psi^{\dagger})_j \D_s x^*\Bigr]
\end{align}
where
\begin{eqnarray}
& z = e^{\phi}\,, \qquad\qquad
z^M = e^{\phi} u^M\,, \qquad\qquad M=1,\dots 6 & \nonumber\\
& \displaystyle u^{a} = \frac{y^{a}}{1+\frac{1}{4}y^2}\,, \qquad\qquad
u^{6} = \frac{1-\frac{1}{4}y^2}{1+\frac{1}{4}y^2}\,, \qquad\qquad y^2\equiv \sum_{a=1}^5 (y^a)^2\,, \quad\qquad a=1,...,5 & \label{eq:redef}
\end{eqnarray}
and $\D_s \equiv \partial_s-1$. The $\rho^{M}_{ij} $ matrices are the off-diagonal blocks of
6d gamma matrices in chiral representation. $(\rho^{MN})_i^{\phantom{i}j} = (\rho^{[M}\rho^{\dagger N]})_i^{\phantom{i}j}$ and $(\rho^{MN})^i_{\phantom{i}j} = ( \rho^{\dagger [M}\rho^{N]})^i_{\phantom{i}j}$ are the $SO(6)$ Lorentz matrices.

The Dirac form \cite{Zarembo:2011ag} is achieved from the action of \cite{Giombi:2010bj}, by packaging the $\eta$ and $\theta$ fermions appearing in the latter into Dirac two-component spinors as follows
\begin{equation}
\psi^i = \left(\begin{array}{c} 
\eta^i \\
(\rho_6^{\dagger})^{ij}\theta_j
\end{array}\right)
\qquad\qquad
\psi^{\dagger}_i = \left( \eta_i, \th^j (\rho^6)_{ji} \right) \qquad\qquad i=1,\dots 4
\end{equation}
The gamma matrices are
\begin{equation}
\g^t = -\sigma_1 \qquad\qquad \g^s = \sigma_3
\end{equation}
and $\bar\psi \equiv \psi^{\dagger}\g^t$, as usual. The projectors appearing in the Lagrangian are defined as $\P_{\pm} \equiv \frac12 \left( \mathbb{1} \pm \g^s \right)$, where $\mathbb{1}$ is the $2\times 2$ identity matrix.

Expanding in the fields at second order 
\begin{equation}
{\cal L}_2  = \partial_{\alpha} \phi\, \partial_{\alpha} \phi +4\,\phi^2 +
\partial_{\alpha} x\, \partial_{\alpha} x^*
+2\, x\, x^{*}
+\partial_{\alpha}y^a\partial_{\alpha}y^a
+ 2\,i\, \bar \psi_i \left(\slashed{\partial} + \mathbb{1} \right)\psi^i
\label{eq:quadratic}
\end{equation}
the spectrum of excitations of the model is inferred, which consists of:
\begin{itemize}
\item a mass $\sqrt{2}$ complex scalar $x$, which together with its complex conjugate represents the insertion of a positive and negative helicity gluon on the GKP vacuum.
\item a mass 2 scalar $\phi$ which from the point of view of the GKP integrable model does not represent an elementary excitation at finite coupling, but is rather interpreted as a composite two-fermion virtual state \cite{Basso:2014koa,Zamolodchikov:2013ama}. The fact that this object is not a proper asymptotic state of the theory renders the computation of matrix elements thereof rather meaningless. Nevertheless, it was argued in \cite{Fioravanti:2015dma} that at strictly infinite coupling the $\phi$ scalars ought to be interpreted as real physical bosons, which were baptised {\it mesons} by the authors. We adopt here this interpretation and nomenclature and compute their S-matrix elements at first order at strong coupling.
\item 5 massless scalars $y^a$, $a=1,\dots 5$, which are the would-be Goldstone bosons originating from spontaneously breaking the original $SO(6)$ invariance of the action to $SO(5)$, which in turn is due to selecting a particular point in $S^5$ for the cusp vacuum. As already clarified in the literature, the $SO(6)$ symmetry is restored by the onset of nonperturbative effects, which consequently provide an exponentially small mass for these scalars. This is captured by the full description of the excitations of the GKP string from integrability, where these scalars represent holes in the GKP vacuum. However these phenomena are not visible in a perturbative approach from the action \eqref{eq:action}.
Moreover the interactions of the massless scalars in \eqref{eq:action} trigger the emergence of IR divergences in loop computations (or even unphysical $1/0$ singularities for amplitudes at tree level) which make the perturbative expansion ill-defined and cast doubts on its validity.
As a consequence, we anticipate that amplitudes involving massless scalars are likely to produce incorrect results. At best the S-matrix elements are just not comparable to those of the ABA approach and violate its underlying $SU(4)$ symmetry, in the worst case scenario they are ill-defined.
We discuss this point further in Section \ref{sec:scalars}.
\item 4 mass 1 Dirac fermions $\psi^i$ ($\psi^{\dagger}_i$), $i=1,\dots 4$, transforming in the $\bf{4}$ ($\bf{\bar 4}$) representation of $SU(4)$, which are mapped to insertions of fermionic excitations on the GKP vacuum.
The fermions are in perfect correspondence with the degrees of freedom of the ABA description. In particular they form multiplets of its $SU(4)$ symmetry.
However, it is clear that the interaction terms in the Lagrangian \eqref{eq:lagrangian} break this symmetry~\footnote{Notice that the Lagrangian \eqref{eq:lagrangian} is $SU(4)$ invariant, however it does not admit a trivial vacuum and one has to break the $SU(4)$ symmetry as in \eqref{eq:redef} to obtain a well defined perturbative expansion.}. This occurs for instance in the coupling with the massless scalars. Therefore one can foresee that problems might occur computing amplitudes of fermions whenever $SU(4)$ breaking interactions undermine the invariance of scattering processes under this expected symmetry.
\end{itemize}
In this paper we analyse the $2\to 2$ tree level scattering of such particles, by computing them with Feynman diagrams.
The Feynman rules are as follows.
From the quadratic action \eqref{eq:quadratic} we derive the propagators
\begin{align}
\langle x(p)x^*(-p) \rangle &= \raisebox{-1mm}{\includegraphics[width=3cm]{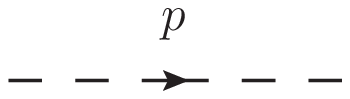}} = \frac{1}{2g}\, \frac{2}{p^2+2}\nonumber\\
\langle \phi(p)\phi(-p) \rangle &= \raisebox{-1mm}{\includegraphics[width=3cm]{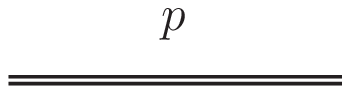}} = \frac{1}{2g}\, \frac{1}{p^2+4}\nonumber\\
\langle y^a(p)y^b(-p) \rangle &= \raisebox{-1mm}{\includegraphics[width=3cm]{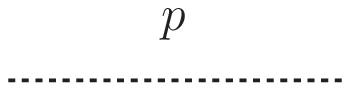}} = \frac{1}{2g}\, \frac{\delta^{ab}}{p^2}\nonumber\\
\langle \psi^i(p)\bar\psi_j(-p) \rangle &= \raisebox{-1mm}{\includegraphics[width=3cm]{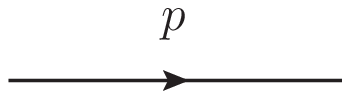}} = \frac{1}{2g}\, i\, \frac{i\slashed{p}-\mathbb{1}}{p^2+1}\, \delta^i_{\phantom{i}j}\nonumber
\end{align}
with the notation we use for drawing Feynman diagrams throughout the article.
Interaction vertices are given by $-\frac12$ those appearing in the Lagrangian giving rise to a consistent expansion in the effective coupling $T$. They are listed in Appendix \ref{app:lagr_exp} for completeness.

We assign momenta $p_1$ and $p_2$ to the incoming scattering particles and $p'_1$ and $p'_2$ to the outgoing ones.
Their components are 
\begin{equation}
p_i = \left( e_i, \ppp_i \right)
\end{equation}
with imaginary energy in the euclidean.
On-shell, we parameterize the momenta of massive particles with hyperbolic rapidities as
\begin{equation}
p_i = m_i \left( i \cosh{\theta_i} , \sinh{\theta_i} \right)
\end{equation}
There are in general two solutions to the momentum conservation constraints with relativistic particles: the first is forward scattering $p'_1=p_1$, $p'_2 = p_2$, the second is backward scattering which for particles of equal mass reads $p'_1=p_2$, $p'_2 = p_1$, and for different masses has a complicated solution.
Integrability predicts that backward scattering should be absent, which is a statement we also want to verify directly.

Solving the momentum conservation $\delta$ functions produces a Jacobian
\begin{equation}
J = \frac{1}{4\, (e_2 \ppp_1 - e_1 \ppp_2)}
\end{equation}
which we have to add to the amplitude.
Fermionic external states yield the polarization Dirac spinors
\begin{equation}
u(p) = \frac{1}{\sqrt{e}} \left(\begin{array}{c}
e \\
\ppp - i
\end{array}\right)
\end{equation}
Since we will not scatter antifermions, $u(p)$ and its conjugate
\begin{equation}
\bar u(p) = \frac{1}{\sqrt{e}} \left( \ppp + i , -e \right)
\end{equation}
are the only polarization spinors needed. Note that the sign of $e$ changes, according to its imaginary nature. The normalization comes in such a way that
\begin{equation}
\bar u(p) u(p) = 2i m = 2i
\end{equation}

The action we use contains an overall factor $\frac{\sqrt{\lambda}}{4\pi} = \frac{T}{2} \equiv g$ \footnote{Note the different convention for the coupling with respect to \cite{Fioravanti:2015dma}, where $g=\frac{\sqrt{\lambda}}{2\sqrt{2}\pi}$.}.
In order to have a standard form for the kinetic terms, we normalize each particle in the initial and final states with a factor $N=1/\sqrt{2g}$, apart from the $x$, $x^*$ scalars whose kinetic terms is off by an extra factor of 2 and are thus normalized with $N_{x}=1/\sqrt{g}$.

Therefore the S-matrix elements read
\begin{equation}\label{eq:S}
S(p_1,p_2) = 1 - \frac{N_1^2 N_2^2}{4\,(e_2\ppp_1-e_1\ppp_2)}\, A(p_1,p_2) + {\cal O}(g^{-2})
\end{equation}
and we compute $A(p_1,p_2)$ with Feynman diagrams.
With the Feynman rules outlined above each interaction vertex has a power of the coupling, whereas propagators introduce an inverse power. Then it is straightforward to see that at tree level $A$ is of order $g$ and therefore $S$ scales as $g^{-1}$.

\section{Scattering of gluons}

\subsection{Same helicity scattering}

We start considering scattering of two transverse gauge excitations of the same helicity $xx\to xx$.
Since the particles are identical, we can restrict to, e.g., the forward solution to the momentum conservation conditions, and sum the diagrams in Figure \ref{fig:treexx}, which correspond to the $t$- and $u$-channel exchange of a mass 2 scalar. 
\FIGURE{
\centering
\includegraphics[width=0.7\textwidth]{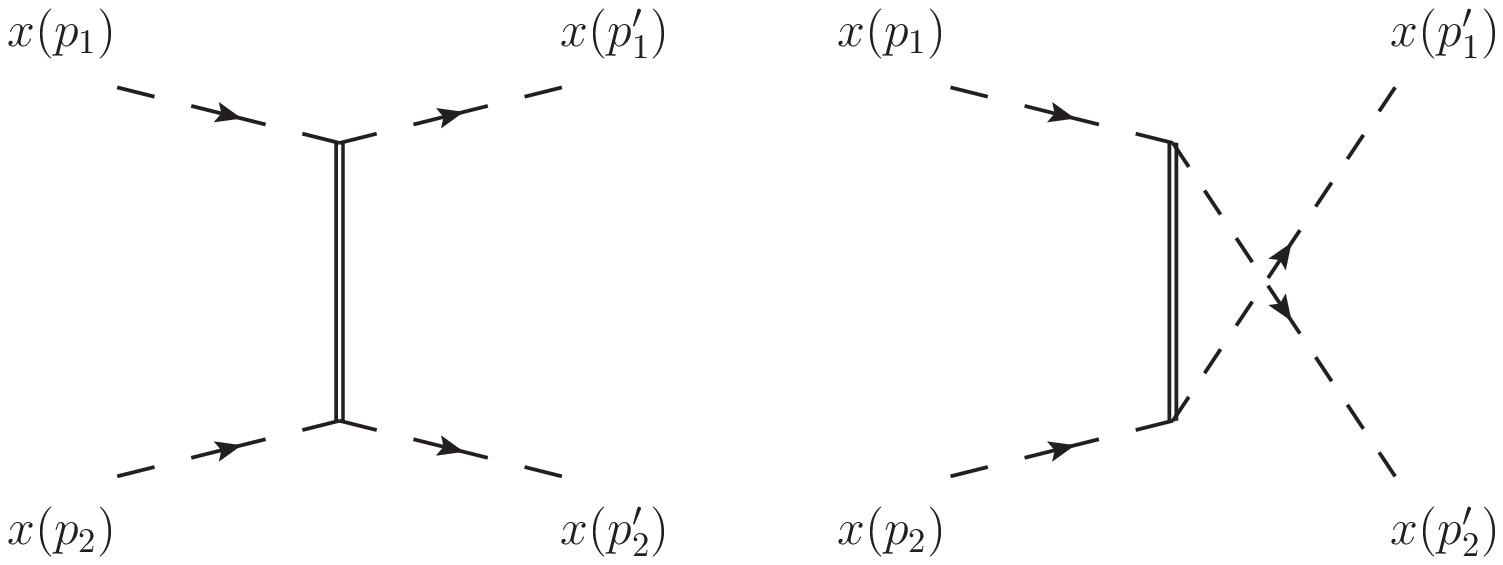}
\caption{Tree-level diagrams for $xx\to xx$ scattering. The exchanged particle is the $\phi$ scalar.}
\label{fig:treexx}
}
Using our euclidean action, the amplitude evaluates 
\begin{equation}\label{eq:treeamplitude}
A^{gg}(p_1,p_2) = 8g\, \left(\ppp_1^2+1\right) \left(\ppp_2^2+1\right) \left( \frac{1}{4} + \frac{1}{(p_1-p_2)^2+4} \right) + {\cal O}(g^0)
\end{equation}
where the two terms in the parenthesis come from the $t$- and $u$-channels of the diagrams in Figure \ref{fig:treexx}, respectively.
Hence the total S-matrix element reads
\begin{equation}
S^{gg}(p_1,p_2) = 1-\frac{2}{g}\, \frac{\left(\ppp_1^2+1\right) \left(\ppp_2^2+1\right)}{4\,(e_2 \ppp_1 - e_1 \ppp_2)}\, \frac{(p_1-p_2)^2+8}{(p_1-p_2)^2+4} + {\cal O}(g^{-2})
\end{equation}
which can be written in terms of hyperbolic rapidities as
\begin{equation}\label{eq:treeamplitude2}
S^{gg}(\th_1,\th_2) = 1 + \frac{i}{g}\, \frac{\cosh{2 \theta_1}\, \cosh{2 \th_2}\, \cosh^2{\frac{\theta_1-\theta_2}{2}}}{\sinh{2 (\theta_1-\theta_2)}} + {\cal O}(g^{-2})
\end{equation}

\subsection{Opposite helicity scattering}

We now turn to the scattering of two transverse gauge excitations with opposite helicity $xx^*\to xx^*$.

The tree-level amplitude is given by the sum of the diagrams in Figure \ref{fig:treexxb}.
\FIGURE{
\centering
\includegraphics[width=0.7\textwidth]{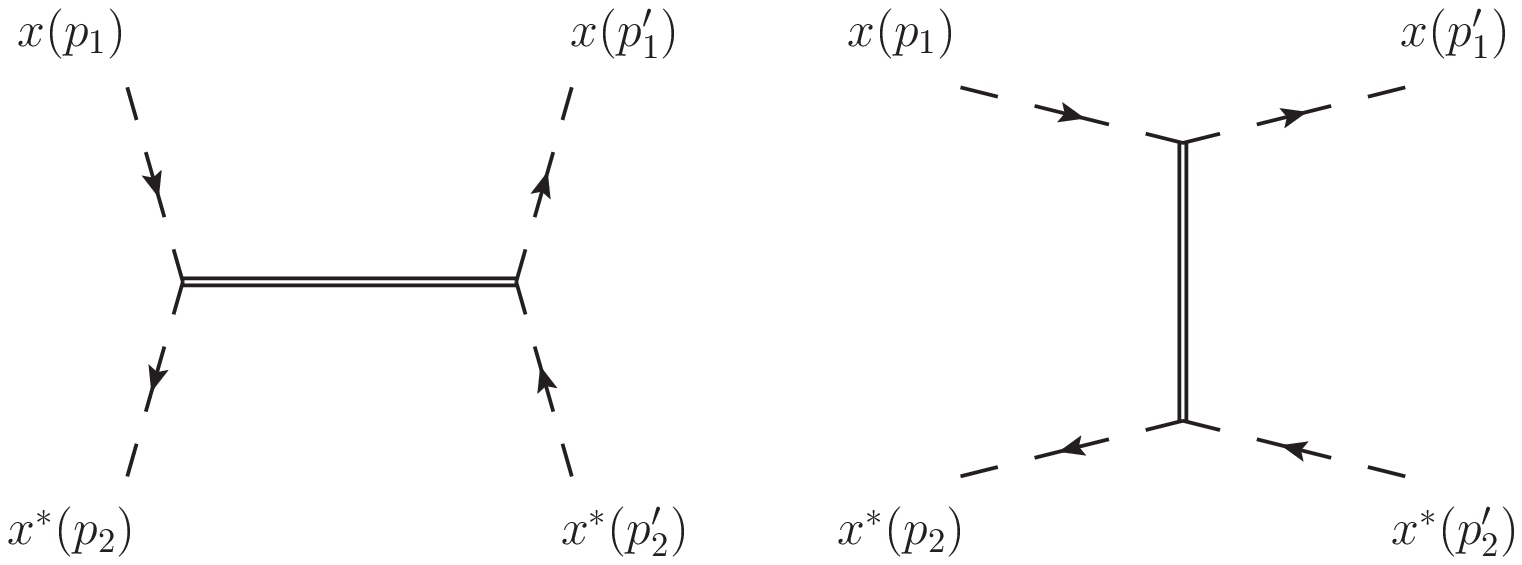}
\caption{Tree-level diagrams for $xx^*\to xx^*$ scattering. The exchanged particle is the $\phi$ scalar.}
\label{fig:treexxb}
}

We begin considering forward scattering, where the particles do not exchange their momenta.
This gives the tree level amplitude
\begin{equation}\label{eq:treexxbforw}
A^{gg^*}(p_1,p_2;p_1,p_2) = 8g\, \left(\ppp_1^2+1\right) \left(\ppp_2^2+1\right) \left( \frac{1}{4} + \frac{1}{(p_1+p_2)^2+4} \right) + {\cal O}(g^0)
\end{equation}
where the notation stresses the forward kinematic configuration.
In hyperbolic rapidities this leads to the expression
\begin{equation}\label{eq:treexxb}
S^{gg^*}(\th_1,\th_2;\th_1,\th_2) = 1+\frac{i}{g}\, \frac{\cosh{2 \theta_1}\, \cosh{2 \th_2}\, \tanh{\frac{\theta_1-\theta_2}{2}} }{\cosh{(\theta_1-\theta_2)}} + {\cal O}(g^{-2})
\end{equation}

In the backward scattering kinematic configuration, interestingly, the two tree-level diagrams of Figure \ref{fig:treexxb} cancel exactly leaving a vanishing result
\begin{equation}\label{eq:treexxbback}
{\cal A}^{gg^*}(p_1,p_2;p_2,p_1) = 8g\, \left(\ppp_1^2+1\right) \left(\ppp_2^2+1\right) \left(\frac{1}{(p_1+p_2)^2+4} + \frac{1}{(p_1-p_2)^2+4}\right) = 0
\end{equation}
where the last equality follows from the identity
\begin{equation}\label{eq:kinid}
(p_1+p_2)^2 + 4 = -(p_1-p_2)^2-4
\end{equation}
which holds for mass $\sqrt{2}$ particles.

\subsection{Comparison to integrability results}

We compare the results obtained for gluon scattering from the string sigma model with the predictions from the ABA.
For the same helicity process, to lowest order in the strong coupling expansion, the integrability result reads
\begin{equation}
S^{gg}(\bar u_1,\bar u_2) = 1 + \frac{1}{2g(\bar u_1-\bar u_2)}\left(1+\frac{1}{2}\left(\frac{1+\bar u_1}{1-\bar u_1}\,\frac{1-\bar u_2}{1+\bar u_2}\right)^{\frac{1}{4}}+\frac{1}{2}\left(\frac{1+\bar u_1}{1-\bar u_1}\,\frac{1-\bar u_2}{1+\bar u_2}\right)^{-\frac{1}{4}}\right) + {\cal O}(g^{-2})
\end{equation}
in terms of (rescaled: $\bar u_i = \frac{u_i}{2g}$) Bethe rapidities, which can be mapped to hyperbolic ones using
\begin{equation}\label{eq:rapidity}
\bar u_i = \tanh 2 \th_i
\end{equation}
to lowest order in the strong perturbative regime.
This gives \cite{Fioravanti:2013eia}
\begin{equation}
S^{gg}(\th_1,\th_2) = 1 + \frac{i}{\sqrt{2}g} \left( \frac{1}{\tanh{2\theta_1}-\tanh{2\theta_2}} + \frac{\cosh{2\theta_1} \cosh{2\theta_2}}{2\sinh{(\theta_1-\theta_2)}}\right) + {\cal O}(g^{-2})
\end{equation}
which coincides with the perturbative result \eqref{eq:treeamplitude2}.

For opposite helicities, the result \eqref{eq:treexxb} for forward kinematics can be directly compared to that quoted in \cite{Fioravanti:2015dma}
\begin{equation}
S^{gg^*}(\th_1,\th_2) = 1 + \frac{i}{\sqrt{2}g} \left( -\frac{1}{\tanh{2\theta_1}-\tanh{2\theta_2}} + \frac{\cosh{2\theta_1} \cosh{2\theta_2}}{2\sinh{(\theta_1-\theta_2)}}\right) + {\cal O}(g^{-2})
\end{equation}
showing agreement.
In addition, we remark that integrability predicts the ratio between the S-matrices for same helicity and opposite helicity in terms of Bethe rapidities \cite{Basso:2013aha}
\begin{equation}
\frac{S^{gg}(u_1,u_2)}{S^{gg^*}(u_1,u_2)} = \frac{u_1-u_2+i}{u_1-u_2-i}
\end{equation}
This statement holds non-trivially at all orders. Rescaling rapidities as $u_i\rightarrow 2g\bar u_i$ and expanding it at first order in perturbation theory for large $g$, we can  appreciate that it has the simple translation in terms of external momenta
\begin{equation}
\frac{S^{gg}(\bar u_1,\bar u_2)}{S^{gg^*}(\bar u_1,\bar u_2)}-1 \propto \frac{8}{g} \frac{1}{(p_1+p_2)^2+4} + {\cal O}(g^{-2})
\end{equation}
which comes precisely from subtracting the dynamical factors of the amplitudes \eqref{eq:treeamplitude} and \eqref{eq:treexxbforw}, using again the kinematic identity \eqref{eq:kinid}.

The integrability results also predict that backward scattering is absent in this process, to all orders. With \eqref{eq:treexxbback} we are able to test this prediction at lowest order in perturbation theory at strong coupling.

\section{Scattering of gluons with other particles}

In this section we compute the amplitudes for scattering of a gluon with a different particle, which might be a fermion, a massless scalar or a meson which, as recalled in the Introduction, we identify with the mass 2 scalar $\phi$ in the spectrum of the string excitations. Anticipating that amplitudes involving the massless scalars are troublesome, we restrict our attention here to scattering of massive excitations only and defer the discussion on $y$ scalars to section \ref{sec:scalars}.

\subsection{Gluon-meson scattering}

This process can be computed through the Feynman diagrams shown in Figure \ref{fig:treexphi}.
\FIGURE{
\centering
\includegraphics[width=0.7\textwidth]{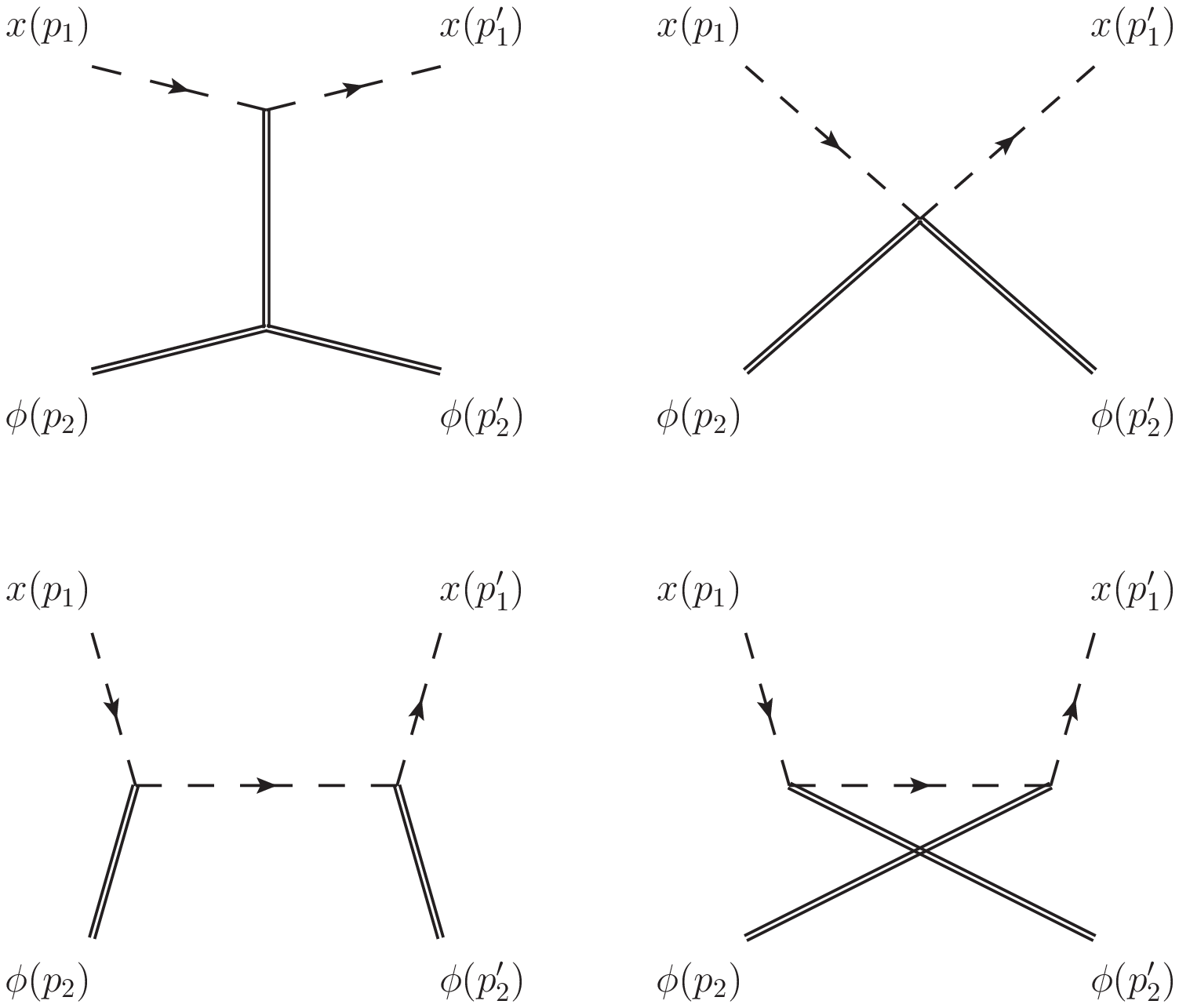}
\caption{Tree-level diagrams for $x\phi\to x\phi$ scattering.}
\label{fig:treexphi}
}

The amplitude evaluates in general
\begin{align}\label{eq:amptreexphi}
A^{gM}(p_1,p_2;p_1',p_2') &= -4\, \frac{(i \ppp_1+1)(-i \ppp_1'+1)}{(p_1-p_1')^2+4}\left(-e_2 e_2'+e_2^2+e_2'^2+\ppp_2 \ppp_2'-\ppp_2^2-\ppp_2'^2\right)+ \nonumber\\&
+ \frac{8\left[(\ppp_1+\ppp_2)^2+1\right]}{(p_1+p_2)^2+2}\, (i \ppp_1+1)(-i \ppp_1'+1)+  \nonumber\\&
+ \frac{8\left[(\ppp_1-\ppp_2')^2+1\right]}{(p_1-p_2')^2+2}\, (i \ppp_1+1)(-i \ppp_1'+1)
-8(i \ppp_1+1)(-i \ppp_1'+1) + {\cal O}(g^{0})
\end{align}
For forward scattering the amplitude takes the form
\begin{equation}\label{eq:treexphi}
A^{gM}(p_1,p_2) = 2 g\, (1+\ppp_1^2) \left(-8-e_2^2+\ppp_2^2 + \frac{8 \left[1+(\ppp_1-\ppp_2)^2\right]}{2+(p_1-p_2)^2} + \frac{8 \left[1+(\ppp_1+\ppp_2)^2\right]}{2+(p_1+p_2)^2}\right) + {\cal O}(g^{0})
\end{equation}
leading to the S-matrix element
\begin{equation}\label{eq:treexphi2}
S^{gM}(\th_1,\th_2) = 1 - \frac{i}{\sqrt{2}\, g} \frac{\cosh{2 \theta_1}\, \sinh{2 \theta_2}\, \cosh{(\theta_1-\theta_2)}}{\cosh{2 (\theta_1-\theta_2)}} + {\cal O}(g^{-2})
\end{equation}
The second solution to the momentum conservation $\delta$ functions has an unpleasant form which produces a nasty expression for the amplitude \eqref{eq:amptreexphi} in this regime. Nevertheless this simplifies to 0, showing that the scattering is reflectionless. 

\subsection{Gluon-fermion}

We turn to scattering between a gluon of positive/negative helicity with a fermion.
We start with the process $\psi x \to \psi x$, whose relevant Feynman diagrams are displayed in Figure \ref{fig:treexpsi}.
\FIGURE{
\centering
\includegraphics[width=0.7\textwidth]{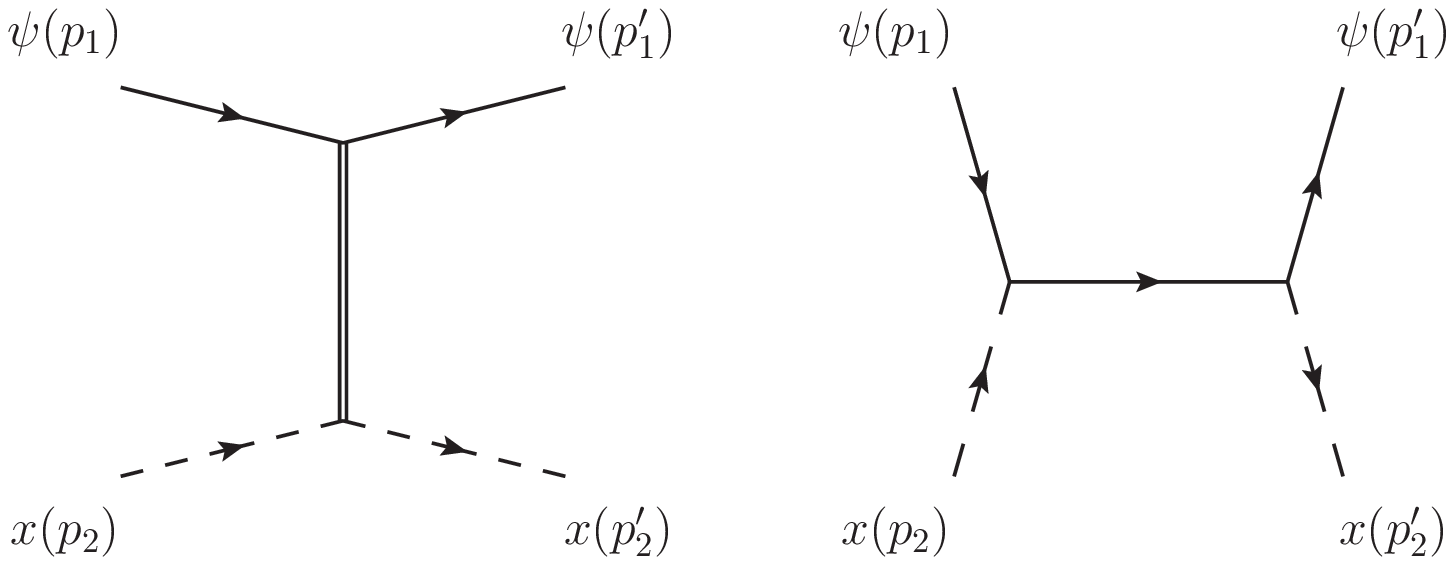}
\caption{Tree-level diagrams for $\psi x \to \psi x$ scattering.}
\label{fig:treexpsi}
}

The algebra of the two diagrams gives (each line comes from a different graph)
\begin{align}\label{eq:amptreexpsi}
A^{fg}(p_1,p_2;p_1',p_2') &= -8 i\, g\, \bar u(p'_1) \left[(-i\ppp'_1-1)\P_{+} + (i\ppp_1-1)\P_{-} \right] u(p_1)\, \frac{(i \ppp_2-1)(-i \ppp'_2-1)}{(p_1-p'_1)^2+4} + \nonumber\\&
-8 i\, g\, \bar u(p'_1)\, \P_{-}\, \frac{i\cancel{(p_1+p_2)}-\mathbb{1}}{(p_1+p_2)^2+1}\, \P_{+}\, u(p_1)\, (i p_2-1)(-i p'_2-1) + {\cal O}(g^{0})
\end{align}
which summed and evaluated for forward kinematics with hyperbolic rapidities gives the simple result
\begin{equation}\label{eq:treexpsi}
S^{fg}(\th_1,\th_2) = 1 - \frac{i}{4\,g}\, \frac{\cosh{2 \theta_2}\, \sinh{2 \theta_1}}{1+\sqrt{2}\, \cosh{(\theta_1-\theta_2)}} + {\cal O}(g^{-2})
\end{equation}
As before, the evaluation of the expression for the amplitude in backward kinematics is more complicated but eventually vanishes.

Considering the process $\psi x^* \to \psi x^*$, we evaluate the Feynamn diagrams of Figure \ref{fig:treexbpsi}.
\FIGURE{
\centering
\includegraphics[width=0.7\textwidth]{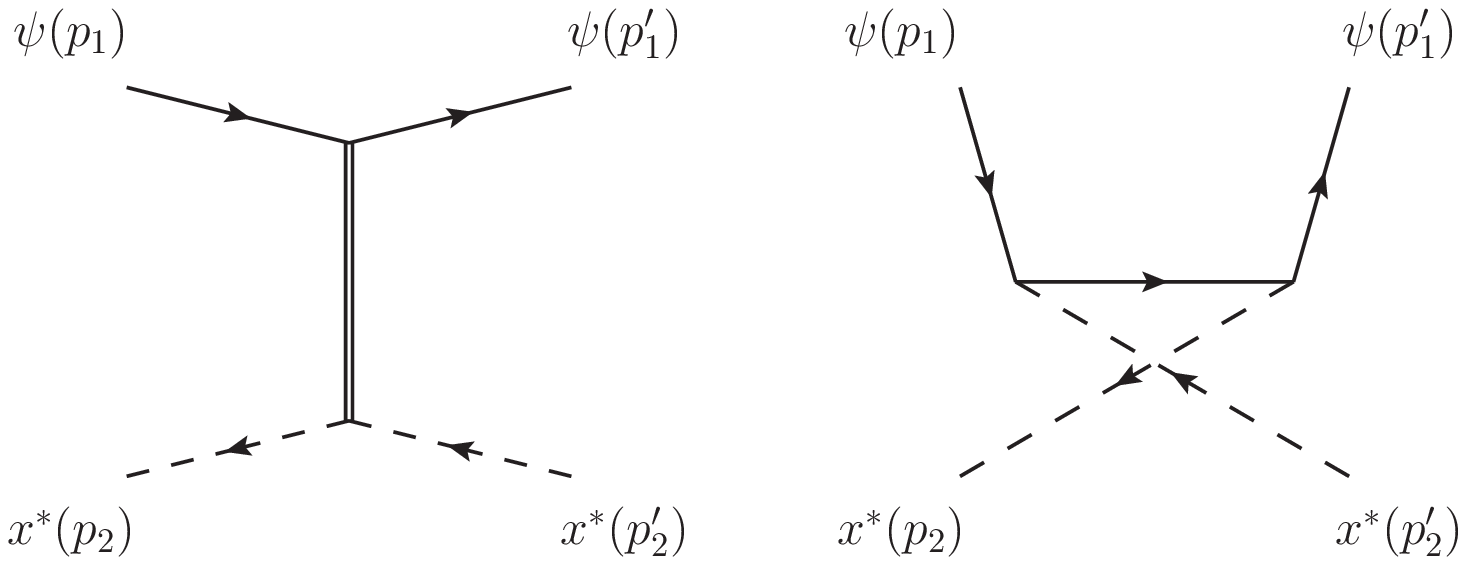}
\caption{Tree-level diagrams for $\psi x^* \to \psi x^*$ scattering.}
\label{fig:treexbpsi}
}
The first gives the same contribution as for the first diagram of Figure \ref{fig:treexpsi}, spelled out in the first line of \eqref{eq:amptreexpsi}, whereas the second differs and is given by the expression
\begin{equation}
-8 i\, g\, \bar u(p'_1)\, \P_{-}\, \frac{i\cancel{(p_1-p'_2)}-\mathbb{1}}{(p_1-p'_2)^2+1}\, \P_{+}\, u(p_1)\, (i p_2-1)(-i p'_2-1)
\end{equation}
The combination of the two terms in forward kinematics is such that only a relative sign changes with respect to the previous result \eqref{eq:treexpsi}
\begin{equation}\label{eq:treexbpsi}
S^{fg}(\th_1,\th_2) = 1 - \frac{i}{4\, g}\, \frac{\cosh{2 \theta_2}\, \sinh{2 \theta_1}}{-1+\sqrt{2}\, \cosh{(\theta_1-\theta_2)}} + {\cal O}(g^{-2})
\end{equation}
Backward scattering is vanishing.

\paragraph{Comparison to integrability}

Following \cite{Fioravanti:2015dma}, the ABA predicts that the meson-gluon scattering phase has the strong coupling expansion
\begin{equation}
S^{gM}(\th_1,\th_2) = 1-\frac{i}{\sqrt{2}\, g}\, \frac{\cosh{(\theta_1-\theta_2)}}{\coth{2 \theta_2} - \tanh{2 \theta_1}} + {\cal O}(g^{-2})
\end{equation}
which is easily seen to be equivalent to our perturbative result \eqref{eq:treexphi2}.

Turning to gluon-fermion scattering, we have to compare our results \eqref{eq:treexpsi} and \eqref{eq:treexbpsi} with the integrability predictions
\begin{align}
S^{fg}(\th_1,\th_2) &= 1 + \frac{i}{4\,g}\, \frac{2\cosh{(\theta_1-\theta_2)}-\sqrt{2}}{\tanh{2\theta_2} - \coth{2\theta_1}} + {\cal O}(g^{-2}) \nonumber\\
S^{fg^*}(\th_1,\th_2) &= 1 + \frac{i}{4\,g}\, \frac{2\cosh{(\theta_1-\theta_2)}+\sqrt{2}}{\tanh{2\theta_2} - \coth{2\theta_1}} + {\cal O}(g^{-2})
\end{align}
which show perfect agreement (upon apparently identifying $x\rightarrow g^*$ and $x^* \rightarrow g$, which is just a matter of conventions).
In addition, we have ascertained that these scattering processes are reflectionless, which is a general feature of integrable scattering matrices involving excitations with different masses.

\section{Scattering of mesons}

\subsection{Meson-meson scattering}

We study the scattering of two mass 2 mesons $\phi$.
The relevant Feynman diagrams, shown in Figure \ref{fig:treephiphi}, evaluate to
\begin{align}\label{eq:amptreephiphi}
A^{\phi\phi}(p_1,p_2) &= 8g\, \left( \frac{(e_1^2+e_2^2-e_1 e_2 - \ppp_1^2-\ppp_2^2+\ppp_1 \ppp_2)^2}{(p_1-p_2)^2+4} + \frac{(e_1^2-\ppp_1^2)(e_2^2-\ppp_2^2)}{4} \right) + \nonumber\\&
+ 8g\, \left( \frac{e_1^2+e_2^2+e_1 e_2 - \ppp_1^2-\ppp_2^2-\ppp_1 \ppp_2}{(p_1+p_2)^2+4} \right) + \nonumber\\& 
- 8g\, \left(4 + p_1^2 + p_2^2 \right) + {\cal O}(g^{0})
\end{align}
where we have already selected, e.g., forward kinematics since the particles are identical. 
In particular, the first contribution arises from the sum of the first and third diagrams which are equal to each other. The last diagram just gives a number, on-shell.
\FIGURE{
\centering
\includegraphics[width=0.7\textwidth]{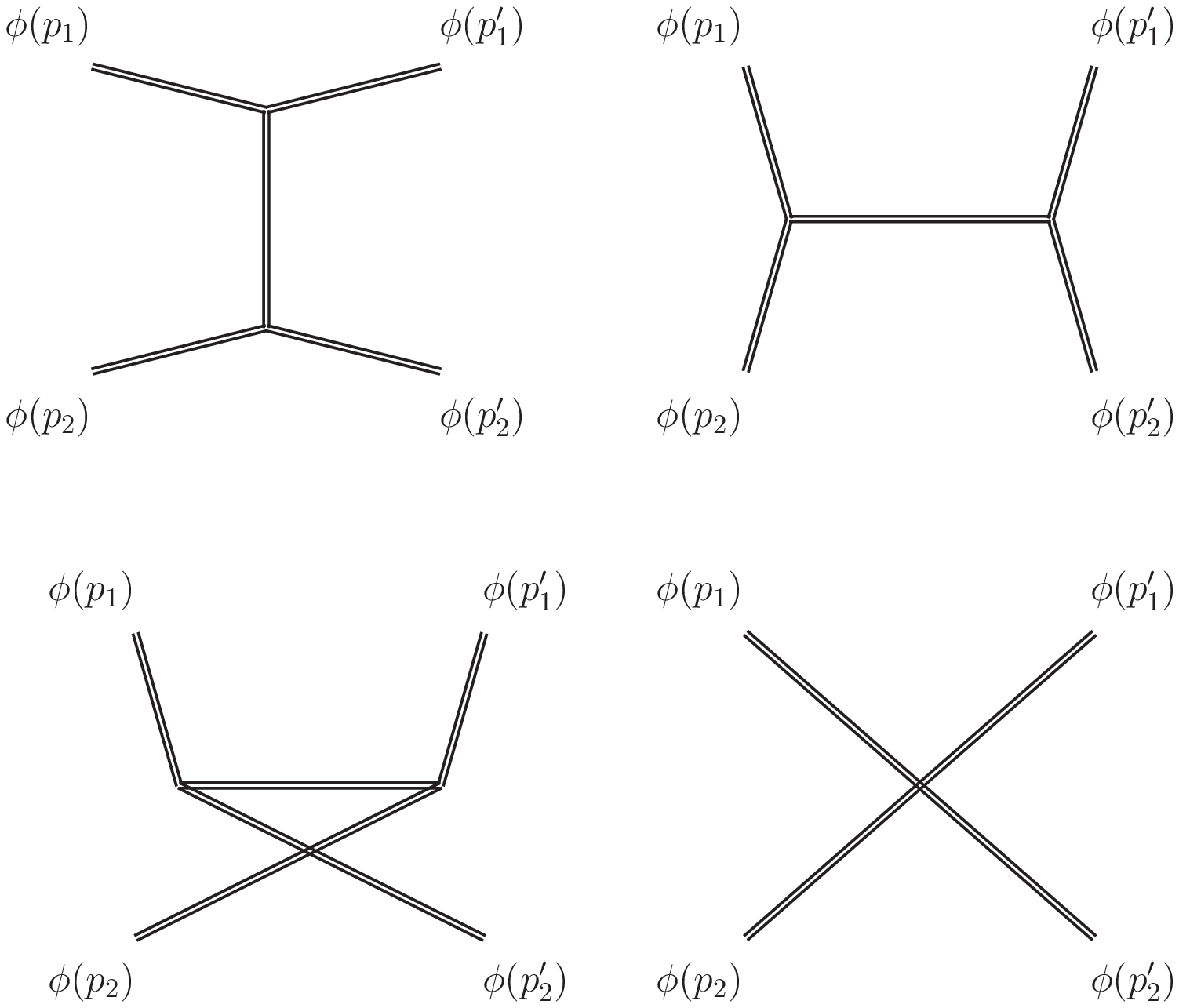}
\caption{Tree-level diagrams for $\phi \phi \to \phi \phi$ scattering.}
\label{fig:treephiphi}
}
Summing them up and turning to hyperbolic rapidities, we obtain the expression
\begin{equation}\label{eq:treephiphi}
S^{MM}(\th_1,\th_2) = 1 + \frac{i}{2\, g}\, \frac{\sinh{2 \theta_1}\, \sinh{2 \theta_2}}{\sinh{(\theta_1 - \theta_2)}} + {\cal O}(g^{-2})
\end{equation}

\subsection{Meson-fermion scattering}

This process involves the diagrams of Figure \ref{fig:treepsiphi}, which read respectively
\begin{align}
A^{fM}(p_1,p_2;p'_1,p'_2) &= 8 i\, g\, \bar u(p'_1) \left[(-i\ppp'_1-1)\P_{+} + (i\ppp_1-1)\P_{-} \right] u(p_1)\times
\nonumber\\&~~~~~~~~~~~~~~~~~ \times \frac{e_2^2+(e'_2)^2-e_2 e'_2 - \ppp_2^2-(\ppp'_2)^2+\ppp_2 \ppp'_2}{(p_1-p'_1)^2+4} + \nonumber\\&
- 8 i\, g\, \bar u(p'_1) \left[(-i\ppp'_1-1)\P_{+} + (i(\ppp_1+\ppp_2)-1)\P_{-} \right] \frac{i\cancel{(p_1+p_2)}-\mathbb{1}}{(p_1+p_2)^2+1}\times
\nonumber\\&~~~~~~~~~~~~~~~~~ \times \left[(-i(\ppp_1+\ppp_2)-1)\P_{+} + (i\ppp_1-1)\P_{-} \right] u(p_1) + \nonumber\\&
- 8 i\, g\, \bar u(p'_1) \left[(-i\ppp'_1-1)\P_{+} + (i(\ppp_1-\ppp'_2)-1)\P_{-} \right] \frac{i\cancel{(p_1-p'_2)}-\mathbb{1}}{(p_1-p'_2)^2+1}\times
\nonumber\\&~~~~~~~~~~~~~~~~~ \times \left[(-i(\ppp_1-\ppp'_2)-1)\P_{+} + (i\ppp_1-1)\P_{-} \right] u(p_1) + \nonumber\\&
+ 8 i\, g\, \bar u(p'_1) \left[(-i\ppp'_1-1)\P_{+} + (i\ppp_1-1)\P_{-} u(p_1) \right] u(p_1) + {\cal O}(g^{0})
\end{align}
\FIGURE{
\centering
\includegraphics[width=0.7\textwidth]{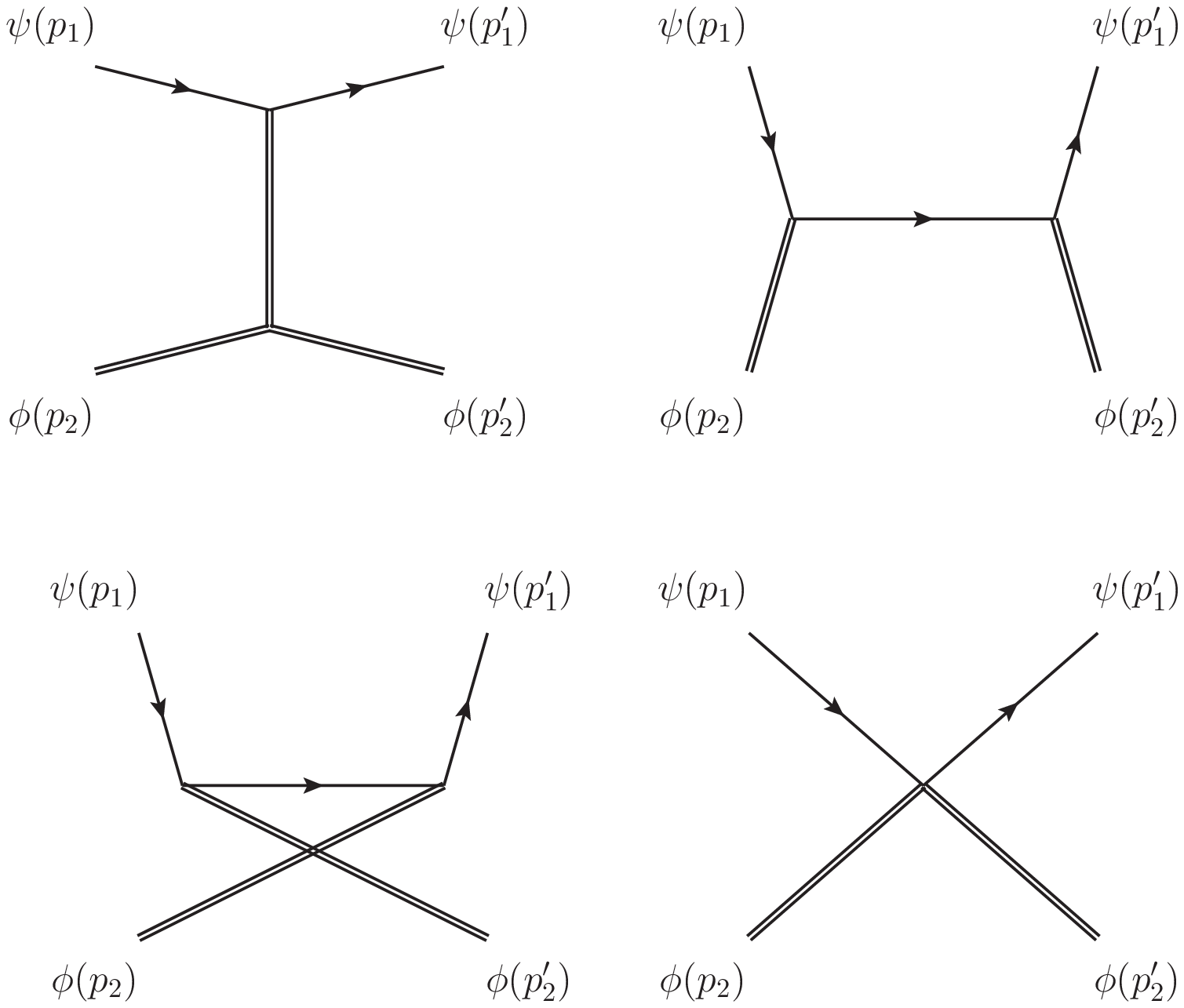}
\caption{Tree-level diagrams for $\psi \phi \to \psi \phi$ scattering.}
\label{fig:treepsiphi}
}
From these we compute the forward scattering phase
\begin{equation}\label{eq:treepsiphi}
S^{fM}(\th_1,\th_2) = 1 + \frac{i}{4g} \frac{\sinh{2 \theta_1} \sinh{2 \theta_2}}{\sinh{(\theta_1 - \theta_2)}} + {\cal O}(g^{-2})
\end{equation}
The solution to backward kinematics generates as usual a cumbersome output, which nevertheless can be shown to vanish.

\subsection{Comparison to integrability results}

The amplitude computed above for meson-meson scattering is found to be in perfect agreement with that quoted in \cite{Fioravanti:2015dma}, formula (C.45).
For fermion-meson scattering the perturbative result also matches the ABA prediction which can be extracted from formulae in section 9 of \cite{Fioravanti:2015dma}, precisely producing \eqref{eq:treepsiphi}.
Again, absence of backward scattering has been verified for these processes at lowest order in perturbation theory.

\section{Amplitudes involving massless scalars}\label{sec:scalars}

We have left aside all amplitudes with scalars as external particles as well as the fermion-fermion scattering, whose tree-level computation involves a massless scalar exchange.
In this section we comment on these processes, which appear problematic to compute using perturbation theory from the action \eqref{eq:lagrangian}, similarly to what was shown to happen for two-point functions \cite{Giombi:2010bj,Zarembo:2011ag,Bianchi:2015laa}.
On the one hand the massless scalars cannot even be identified with the degrees of freedom of the integrable model describing the GKP string as their number differs. Hence it would be quite meaningless to compare their scattering matrices.
On the other hand the massless scalars can cause problems even when they do not appear as external states, but as exchanged particles. This happens for instance when trying to compute fermion-fermion scattering. The massless scalars introduce interactions which break the $SU(4)$ symmetry of the Lagrangian and hence produce a violation of the $SU(4)$ structure expected for fermion-fermion scattering.
Moreover, if treated as massless, an exchange of $y$ scalars in the $t$-channel is plagued by an unphysical $1/0$ singular term caused by the propagator, which signals an inconsistency of the perturbative approach. Finally, the exponentially suppressed mass gap of the theory combined with the logarithimic dependence on the IR cutoff appearing in IR divergent higher loops contributions would invalidate the perturbative result even at tree level \cite{Zarembo:2011ag}.
We verify and address these issues, where possible, studying the aforementioned amplitudes.

\subsection{Fermion-fermion scattering}

First we tackle the amplitude between a pair of fermions.
These particles transform in the $\bf{4}$ representation of $SU(4)$, hence the $2\to 2$ amplitude is a 4-indices tensor of $SU(4)$. Following \cite{Basso:2014koa} we define it as
\begin{equation}\label{eq:fermionamplitude}
\big| \psi^i(p_1) \psi^j(p_2) \big\rangle = S^{ff}(p_1,p_2)^{ij}_{kl} \big| \psi^l(p_2) \psi^k(p_1) \big\rangle
\end{equation}
One could also consider the fermion-antifermion amplitude, but its computation involves a higher number of Feynman diagrams, therefore we focus on \eqref{eq:fermionamplitude} and evaluate the relevant graphs of Figure \ref{fig:treepsipsi}.
\FIGURE{
\centering
\includegraphics[width=1.\textwidth]{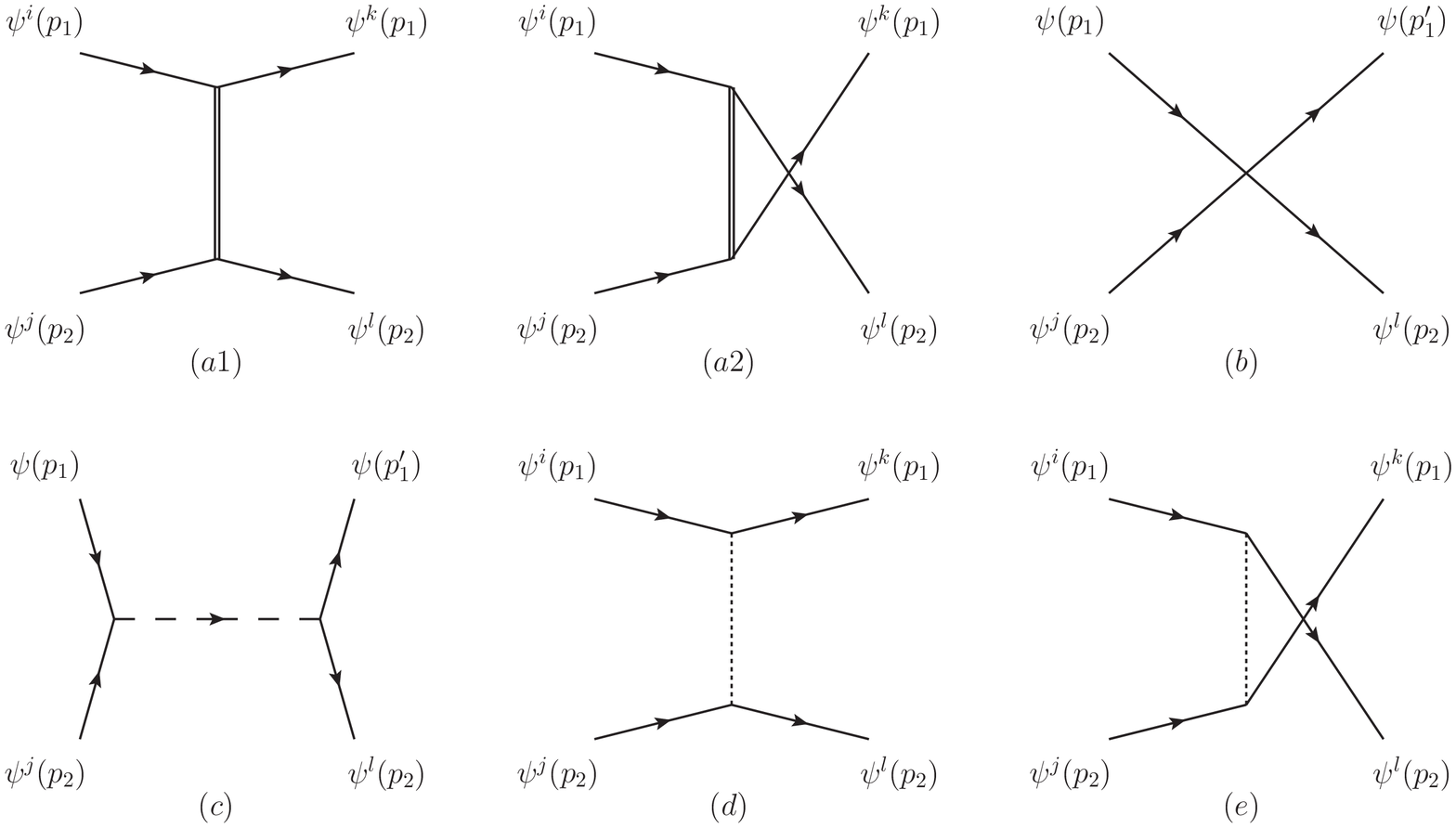}
\caption{Tree-level diagrams for $\psi_i \psi_j \to \psi_k \psi_l$ scattering.}
\label{fig:treepsipsi}
}
The computation of the first four are straightforward and yield separately
\begin{align}\label{eq:fermioncontr}
(a1)&= 8 g \cosh^2{\theta_1} \cosh^2{\theta_2}\, \delta^i_k\delta^j_l \equiv a_1\, \delta^i_k\delta^j_l\nonumber\\
(a2)&= -8 g \cosh{\theta_1} \cosh{\theta_2} \cosh^2{\frac{\theta_1+\theta_2}{2}}\, \delta^i_l\delta^j_k \equiv a_2\, \delta^i_l\delta^j_k \nonumber\\
(b)&= -2 g \cosh{\theta_1} \cosh{\theta_2}\, \left(\delta^i_k\delta^j_l-\delta^i_l\delta^j_k + (\rho^{a6})^i_{\phantom{i}k}(\rho^{a6})^j_{\phantom{j}l}-(\rho^{a6})^i_{\phantom{i}l}(\rho^{a6})^j_{\phantom{j}k}\right) \equiv \nonumber\\& \equiv b \left(\delta^i_k\delta^j_l-\delta^i_l\delta^j_k + (\rho^{a6})^i_{\phantom{i}k}(\rho^{a6})^j_{\phantom{j}l}-(\rho^{a6})^i_{\phantom{i}l}(\rho^{a6})^j_{\phantom{j}k}\right) \nonumber\\
(c)&= 8 g \cosh{\theta_1} \cosh{\theta_2} \left(\cosh{(\theta_1+\theta_2)} + \frac{2\sinh{\theta_1} \sinh{\theta_2}}{\cosh{(\theta_1-\theta_2)}}\right)\, (\rho^6)^{ij}(\rho^\dagger_6)_{kl} \equiv c\, (\rho^6)^{ij}(\rho^\dagger_6)_{kl}
\end{align}
in terms of hyperbolic rapidities.
We note that the diagrams contribute to different tensor structures. In particular, those with a mass 2 meson exchange are proportional to $\delta^i_k\delta^j_l$ and $\delta^i_l\delta^j_k$, respectively, that triggered by a gluon exchange is proportional to $(\rho^6)^{ij}(\rho^\dagger_6)_{kl}$ and the quartic vertex diagram is proportional to $\delta^i_k\delta^j_l-\delta^i_l\delta^j_k$ and $(\rho^{a6})^i_{\phantom{i}k}(\rho^{a6})^j_{\phantom{j}l}-(\rho^{a6})^i_{\phantom{i}l}(\rho^{a6})^j_{\phantom{j}k}$.
The diagrams featuring a massless scalar exchange remain to be evaluated. 
The first is proportional to the tensor structure $(\rho^{a6})^i_{\phantom{i}k}(\rho^{a6})^j_{\phantom{j}l}$ and its algebra is troublesome: momentum conservation in two-dimensional kinematics forces the internal propagator to be singular. This unphysical phenomenon signals that something wrong is happening in the perturbative expansion. One may regulate the propagator with a small mass, which sounds reasonable since the scalars acquire a small nonperturbative mass, after all. With such a regulator the diagram is found to vanish, on-shell, since the numerator is proportional to the fermion on-shell condition.
The diagram with a scalar exchange in the $u$-channel, which contributes to the $(\rho^{a6})^i_{\phantom{i}l}(\rho^{a6})^j_{\phantom{j}k}$ structure, is not singular but vanishes on-shell as well.
The result of such a naive computation is certainly far from the prediction of integrability.
In particular the tensor structure of the result is violating the expected $SU(4)$ symmetry of the integrable model. 
The tensor structures appearing in it are not independent, on the contrary they are related by the tensor identities
\begin{align}\label{eq:tensoridentities}
& (\rho^{a6})^i_{\phantom{i}k}(\rho^{a6})^j_{\phantom{j}l}-(\rho^{a6})^i_{\phantom{i}l}(\rho^{a6})^j_{\phantom{j}k} - 3\, (\delta^i_k\delta^j_l-\delta^i_l\delta^j_k) + 4\, (\rho^6)^{ij}(\rho^\dagger_6)_{kl} = 0\nonumber\\
& (\rho^{a6})^i_{\phantom{i}k}(\rho^{a6})^j_{\phantom{j}l}-(\delta^i_k\delta^j_l-2\delta^i_l\delta^j_k)-2\, (\rho^6)^{ij}(\rho^\dagger_6)_{kl} = 0
\end{align}
Still, if one tries, e.g., to eliminate the $\rho^{a6}$ tensors from the result, it is clear from the very different expressions of the contributions, that there is no chance the  $\rho^6 \rho^\dagger_6$ piece cancels, which would leave $SU(4)$ invariant tensors only.
At this point we conclude that the perturbative approach fails to compute this amplitude and blame the massless scalars for this, along the lines of \cite{Zarembo:2011ag}.
Nevertheless, we can still try to make use of the computation of the diagrams $(a1)$, $(a2)$, $(b)$ and $(c)$ in Figure \ref{fig:treepsipsi}, which looks legitimate, with some experimental physics.
Let's say that the interactions between massless scalars and fermions are not suitable for this computation because of the onset of nonperturbative phenomena which are not accessible via our analysis. As explained in \cite{Zarembo:2011ag}, the massless scalars cause infrared divergences in loop computations, which can be thought of as logarithms of their exponentially small mass. Therefore these logarithms produce positive powers of the coupling, mixing perturbative orders and invalidating perturbation theory. We can imagine that an infinite tower of leading logarithms can be resummed and produce a nonvanishing contribution to the tree level result for the fermion amplitude.
We can also {\it suppose} that the tensor structure of this contribution is proportional to the tree level structures $(\rho^{a6})^i_{\phantom{i}k}(\rho^{a6})^j_{\phantom{j}l}$ and $(\rho^{a6})^i_{\phantom{i}l}(\rho^{a6})^j_{\phantom{j}k}$, thought we admittedly do not have any solid argument to justify this. To parameterize our ignorance on the form of these interactions we introduce the two undetermined functions $x$ and $y$ as order $g$ coefficients of the $\rho^{a6}$ tensors
\begin{equation}\label{eq:parameterization}
x\, (\rho^{a6})^i_{\phantom{i}k}(\rho^{a6})^j_{\phantom{j}l} + y\, (\rho^{a6})^i_{\phantom{i}l}(\rho^{a6})^j_{\phantom{j}k}
\end{equation}
Next we {\it assume} that the scattering process occurs in an $SU(4)$ invariant and integrable fashion and borrow the general expression for such an S-matrix \cite{Berg:1977dp,Basso:2014koa}
\begin{equation}\label{eq:tensorstructure}
S^{ff}(u_1,u_2)^{ij}_{kl} = S^{ff}(u_1,u_2) \left(\frac{u_1-u_2}{u_1-u_2-i}\, \delta^i_k\delta^j_l - \frac{i}{u_1-u_2-i}\, \delta^i_l\delta^j_k \right)
\end{equation}
in terms of Bethe rapidities. The scalar factor $S^{ff}(u_1,u_2)$ encloses the dynamics of the particular integrable model, that is the GKP string in the case at hand.
This assumption is putting some extra crucial ingredient at this point, but let us go ahead with this working hypothesis and see if we get some mileage.
First we expand \eqref{eq:tensorstructure} at strong coupling by first rescaling the Bethe rapidities $u_i=2g\bar u_i$, expanding to first order at $g\to\infty$ and mapping the Bethe rapidities to hyperbolic, $\bar u_i = \coth 2\th_i$ for fermions. This gives
\begin{align}\label{eq:tensorstructurehyp}
S^{ff}(\th_1,\th_2)^{ij}_{kl} &= \left( 1 + \frac{1}{g}\, S^{ff}(\th_1,\th_2)^{(1)} + {\cal O}(g^{-2}) \right) \times\nonumber\\&
\times\left[\left(1+\frac{i}{2\,g}\,\frac{1}{\coth{2\theta_1} - \coth{2\theta_2}}\right) \delta^i_k\delta^j_l -\frac{i}{2\,g}\,\frac{1}{\coth{2\theta_1} - \coth{2\theta_2}}\, \delta^i_l\delta^j_k + {\cal O}(g^{-2}) \right]
\end{align}
On the other hand, using \eqref{eq:parameterization} and \eqref{eq:fermioncontr}, the amplitude reads
\begin{align}
S^{ff}(\th_1,\th_2)^{ij}_{kl} &= 1+\frac{i}{16\,g^2\, \sinh{(\th_1-\th_2)}}\left( (a_1+b)\,\delta^i_k\delta^j_l + (a_2-b)\,\delta^i_l\delta^j_k + c\, (\rho^6)^{ij}(\rho^\dagger_6)_{kl} \right) + \nonumber\\& + x\, (\rho^{a6})^i_{\phantom{i}k}(\rho^{a6})^j_{\phantom{j}l} + y\, (\rho^{a6})^i_{\phantom{i}l}(\rho^{a6})^j_{\phantom{j}k} + {\cal O}(g^{-2})
\end{align}
If we insists that it has to respect the form \eqref{eq:tensorstructurehyp}, we can plug \eqref{eq:tensoridentities} into the equation above in order to eliminate the $\rho^{6a}$ structure and impose that the $\rho^6\rho^\dagger_6$ tensors also drop out.
This leaves us with a linear system in three unknowns, where that we are aiming at is the scalar factor $S^{ff}(\th_1,\th_2)$
\begin{equation}
\left\{\begin{array}{l}\displaystyle
a_1 + b + x - 2y = \frac{i}{2\,g}\,\frac{1}{\coth{2\theta_1} - \coth{2\theta_2}} + \frac{1}{g}\, S^{ff}(\th_1,\th_2)^{(1)}\\\displaystyle
a_2 - b - 2x + y = -\frac{i}{2\,g}\,\frac{1}{\coth{2\theta_1} - \coth{2\theta_2}}\\
c + 2x - 2y = 0
\end{array}\right.
\end{equation}
Solving the system we obtain
\begin{equation}
S^{ff}(\th_1,\th_2) = 1+\frac{i}{4\,g}\,\frac{\cosh{(\th_1-\th_2)}-1}{\coth{2\theta_1} - \coth{2\theta_2}} + {\cal O}(g^{-2})
\end{equation}
which is in precise agreement with the prediction of \cite{Fioravanti:2015dma}.
We want to stress that the derivation above is highly speculative and already assumes integrability as an input.
Still, we find interesting that the perturbative computation of a subset of {\it safe} graphs is able to reproduce the correct result of the fermion scalar factor, which arises from the complicated nonperturbative dynamics of the GKP string.

\subsection{Scattering of massless scalars}
 
We turn to scattering involving massless scalars as external particles.
As mentioned above, although it is possible to construct Feynman diagrams for them starting from the action \eqref{eq:lagrangian}, it is not clear what to compare the objects computed this way to. Indeed the five massless scalars present in the model are not directly mapped to the holes of the integrable GKP string model and the dynamics of the latter is highly nonperturbative.
For instance, from the point of view of the string sigma model \eqref{eq:lagrangian}, the scattering amplitude of a scalar off a gluon vanishes identically at tree level, since there are simply no interaction vertices to construct it.
On the other hand integrability predicts that the amplitude is finite and possesses a contribution of order $g^{-1}$.
Clearly there is a clash between the two approaches.
For other processes there are in principle Feynman diagrams one can construct, but we are skeptical on the possibility of extracting any interesting information from them, given the known shortcomings of the model when addressing quantities that are not $SU(4)$ invariant.

\section{Particle production and factorization}

In this section we provide evidence for the absence of particle production and the factorization of the $3\to3$ particle S-matrix in terms of two-body ones \cite{Zamolodchikov:1978xm}. Let us first recall which structure the factorization constraint assumes when expanded perturbatively. We start from the basic factorization equation
\begin{equation}\label{eq:fact}
S_{123} = S_{12}\, S_{13}\, S_{23}
\end{equation}
where the operators act on a three-particle state and the indices label the scattering particles. In this notation the product of S-matrices is not commutative and the consistency of factorization is provided by the Yang-Baxter equation
\begin{equation}
S_{12}\, S_{13}\, S_{23} = S_{23}\, S_{13}\, S_{12}
\end{equation}
Expanding \eqref{eq:fact} perturbatively as $S = \mathbb{1} - \frac{1}{g}\, T^{(0)} + \mathcal{O}(g^{-2})$ one obtains the tree-level identity
\begin{equation}\label{eq:YB}
T^{(0)}_{123}= T_{12}^{(0)} T_{13}^{(0)} + T_{12}^{(0)} T_{23}^{(0)} + T_{13}^{(0)} T_{23}^{(0)} 
\end{equation}
In the following we show that this identity holds for the $3\to 3$ scattering processes involving bosonic GKP massive excitations.

\subsection{Scattering of three gluons}

Let us start from the simplest case, i.e. the $xxx\to xxx$ S-matrix. 
\FIGURE{
\centering
\includegraphics[width=\textwidth]{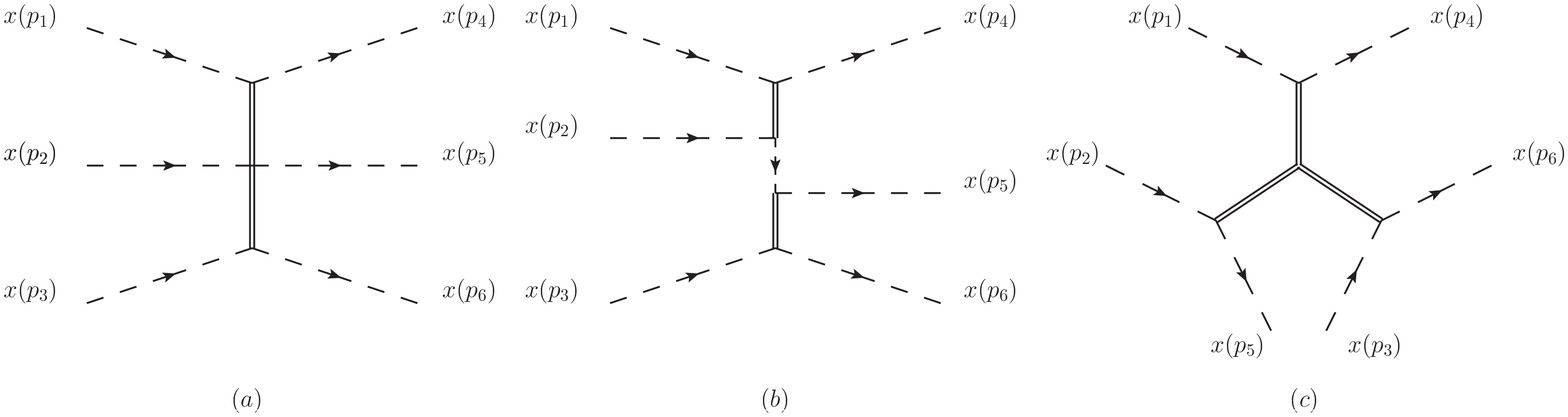}
\caption{Tree-level diagrams for $xxx \to xxx$ scattering.}
\label{fig:treexxx}
}
The contributing diagrams are shown in Figure \ref{fig:treexxx}, with all possible permutations of external momenta. These contributions, with the choice of momenta in the figure and using the shorthand notation $p_{ij}\equiv p_i - p_j$, evaluate to 
\begin{align}
d_1^{xxx} &= - 64g\, \frac{(i \ppp_1-1)(i \ppp_2-1)(i \ppp_3-1)(-i \ppp_4-1)(-i \ppp_5-1)(-i \ppp_6-1)}{\left[p_{14}^2+4\right]\left[p_{36}^2+4\right]} \nonumber\\
d_2^{xxx} &= 64g\, \frac{(i \ppp_1-1)(i \ppp_2-1)(i \ppp_3-1)(-i \ppp_4-1)(-i \ppp_5-1)(-i \ppp_6-1)}{\left[p_{14}^2+4\right]\left[(p_{14}+p_2)^2+2\right]\left[p_{36}^2+4\right]} \left( (\ppp_{14}+\ppp_2)^2 + 1 \right) \nonumber\\
d_3^{xxx} &= 32g\, \frac{\left[ -e_{14} e_{25} - e_{25} e_{36} - e_{36} e_{14} - (e \leftrightarrow \ppp)\right]}{\left[p_{14}^2+4\right]\left[p_{25}^2+4\right]\left[p_{36}^2+4\right]}  \times\nonumber\\& ~~~~ \times
(i \ppp_1-1)(i \ppp_2-1)(i \ppp_3-1)(-i \ppp_4-1)(-i \ppp_5-1)(-i \ppp_6-1) 
\end{align}
The total amplitude is given by the sum of the diagrams above, summed over the 36 permutations of the incoming and outgoing external momenta separately and weighted by the following symmetry factors
\begin{equation}
A^{xxx} \propto \frac12\, d_1^{xxx} + d_2^{xxx} + \frac16\, d_3^{xxx} + \mathrm{perms} = 0
\end{equation} 
and is found to vanish for generic kinematics.
\FIGURE{
\centering
\includegraphics[width=\textwidth]{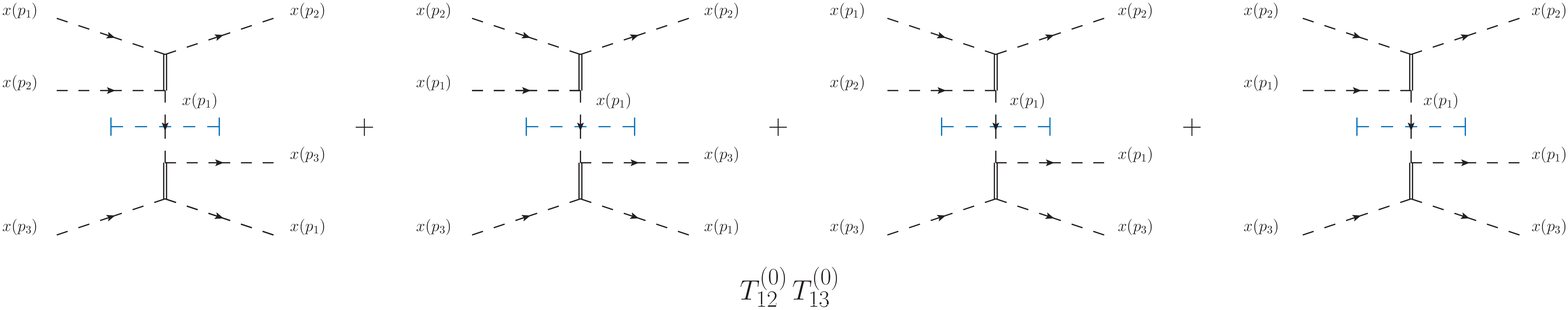}
\caption{Diagrammatic interpretation of the S-matrix factorization for the scattering of three gluons. A similar picture can be derived also for the internal propagator with momentum $p_2$ and $p_3$. The blue dashed line indicates that the propagator has to be replaced by an on-shell $\delta$ function $2i\p \delta(p_1^2-2)$.}
\label{fig:factex}
}
Care has to be taken for special kinematics, for instance whenever $p_1 = p_4$. This automatically forces the other momenta to be equal pairwise, namely $p_2 = p_5$, $p_3=p_6$ or $p_2=p_6$, $p_3=p_5$. In such a situation, and all permutations thereof, the first diagram develops a singularity because of the on-shell intermediate $x$ propagator.
The other diagrams are regular since they do not possess any propagators going on-shell.
With the Feynman prescription the singular propagator splits as usual into a finite, principal value, part and a $\delta$ function. The finite part cancels among the three diagrams as in the non-singular case, whereas the $\delta$ function part produces the only non-vanishing contribution. In Figure \ref{fig:factex} we provide an example of such a situation with the blue dashed line indicating a cut propagator, i.e. an on-shell $\delta$ function. The four singular configurations involving an on-shell propagator with momentum $p_1$ group themselves in such a way that they can be explicitly interpreted as the product of the $t$- and $u$-channel contributions in Figure \ref{fig:treexxb} for the tree-level S-matrices $T^{xx}(p_1,p_2)$ and $T^{xx}(p_1,p_3)$. A similar picture arises for internal propagators with momenta $p_2$ and $p_3$ leading to a factorization of the form\begin{equation}
T^{xxx}(p_1,p_2,p_3) = T^{xx}(p_1,p_2)T^{xx}(p_1,p_3)+T^{xx}(p_1,p_2)T^{xx}(p_2,p_3)+T^{xx}(p_1,p_3)T^{xx}(p_2,p_3)
\end{equation} 
predicted by the Yang-Baxter equation \eqref{eq:YB}.

\subsection{Scattering of three mesons}

A slightly more involved computation can be carried out to ascertain factorization for the $3\to 3$ scattering of mesons.
There are seven relevant topologies of Feynman diagrams contributing to this process, drawn in Figure \ref{fig:treephiphiphi}, with all possible permutations of external legs.
\FIGURE{
\centering
\includegraphics[width=\textwidth]{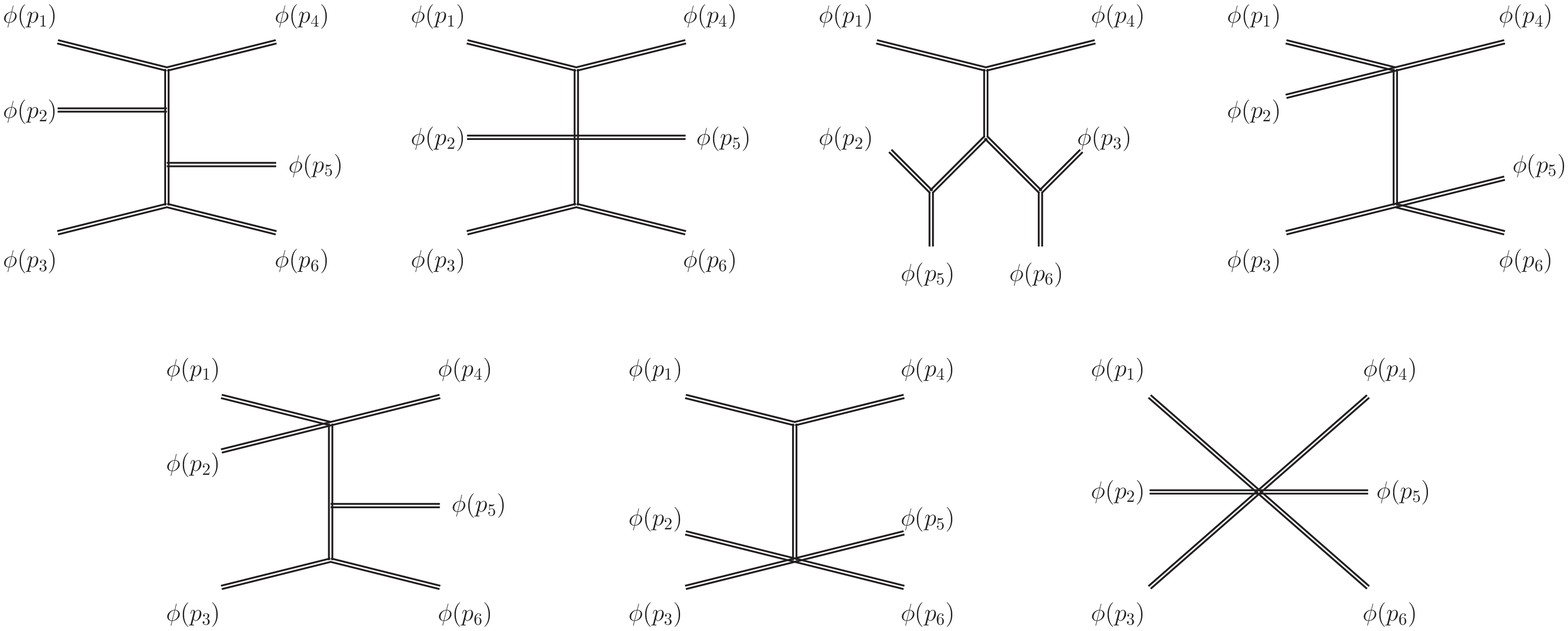}
\caption{Tree-level diagrams for $\phi\phi\phi \to \phi\phi\phi$ scattering.}
\label{fig:treephiphiphi}
}
For the choice of external momenta shown in the picture the diagrams read, using the shorthand notation $p_{ij}\equiv p_i - p_j$
\begin{align}
d_1^{\phi\phi\phi} & = 32g\, \frac{\left[ e_1^2 + e_4^2 - e_1 e_4 - (e \leftrightarrow \ppp) \right]}{\left[p_{14}^2+4\right]\left[(p_{14}+p_2)^2+4\right]\left[p_{36}^2+4\right]}  \left[ e_3^2 + e_6^2 - e_3 e_6 - (e \leftrightarrow \ppp) \right] \times\nonumber\\& ~~~~ \times
\left[ e_2^2 + e_{14}^2 + e_2 e_{14} - (e \leftrightarrow \ppp)\right] 
\left[ e_5^2 + e_{36}^2 - e_5 e_{36} - (e \leftrightarrow \ppp)\right]
\nonumber\\
d_2^{\phi\phi\phi} & = -32g\, \frac{\left[ e_1^2 + e_4^2 - e_1 e_4 - (e \leftrightarrow \ppp) \right]}{\left[p_{14}^2+4\right]\left[p_{36}^2+4\right]} \left[ e_3^2 + e_6^2 - e_3 e_6 - (e \leftrightarrow \ppp) \right] \times\nonumber\\& ~~~~ \times
\left[ 4 - p_2\cdot p_{14} - p_{14}\cdot p_{36} + p_5\cdot p_{14} - p_2\cdot p_{36} + p_5\cdot p_{36} + p_2\cdot p_5 \right] \nonumber\\
d_3^{\phi\phi\phi} & = 32g\, \frac{\left[ e_1^2 + e_4^2 - e_1 e_4 - (e \leftrightarrow \ppp) \right]}{\left[p_{14}^2+4\right]\left[p_{25}^2+4\right]\left[p_{36}^2+4\right]}  \left[ e_3^2 + e_6^2 - e_3 e_6 - (e \leftrightarrow \ppp) \right] \times\nonumber\\& ~~~~ \times
\left[ e_2^2 + e_5^2 - e_2 e_5 - (e \leftrightarrow \ppp)\right] 
\left[ -e_{14} e_{25} - e_{25} e_{36} - e_{36} e_{14} - (e \leftrightarrow \ppp)\right] \nonumber\\
d_4^{\phi\phi\phi} & = -32g\, \frac{\left[ e_3^2 + e_6^2 - e_3 e_6 - (e \leftrightarrow \ppp) \right]}{\left[(p_{14}+p_2)^2+4\right]\left[p_{36}^2+4\right]} \left[ e_5^2 + e_{36}^2 - e_5 e_{36} - (e \leftrightarrow \ppp)\right] \times\nonumber\\& \times
\left[ 4 - p_1\cdot (p_{14}+p_2) - p_2\cdot (p_{14}+p_2) + p_4\cdot (p_{14}+p_2) - p_1\cdot p_2 + p_2\cdot p_4 + p_4\cdot p_1 \right]
\nonumber\\ 
d_5^{\phi\phi\phi} & = 32g\, \frac{1}{\left[(p_{14}+p_2)^2+4\right]} \times\nonumber\\& \times
\left[ 4 - p_1\cdot (p_{14}+p_2) - p_2\cdot (p_{14}+p_2) + p_4\cdot (p_{14}+p_2) - p_1\cdot p_2 + p_2\cdot p_4 + p_4\cdot p_1 \right] \times\nonumber\\& \times
\left[ 4 - p_3\cdot (p_{14}+p_2) + p_6\cdot (p_{14}+p_2) + p_5\cdot (p_{14}+p_2) + p_3\cdot p_6 - p_6\cdot p_5 + p_5\cdot p_3 \right] \nonumber\\
d_6^{\phi\phi\phi} & = 32g\, \frac{\left[ e_1^2 + e_4^2 - e_1 e_4 - (e \leftrightarrow \ppp) \right]}{\left[p_{14}^2+4\right]} \times\nonumber\\& \times
\left[ e_5 e_{14} + e_6 e_{14} - e_2 e_{14} - e_3 e_{14} - e_2 e_3 + e_3 e_6 - e_6 e_5 + e_5 e_3 - (e \leftrightarrow \ppp) \right] \nonumber\\
d_7^{\phi\phi\phi} & = -32g\, \left[ 4 - p_1\cdot p_2 - p_1\cdot p_3 + p_1\cdot p_4 + p_1\cdot p_5 + p_1\cdot p_6 - p_2\cdot p_3 + p_2\cdot p_4 + p_2\cdot p_5 + \right.\nonumber\\&\left. ~~~~ + p_2\cdot p_6 +p_2\cdot p_4 + p_3\cdot p_5 + p_3\cdot p_6 - p_4\cdot p_5 - p_4\cdot p_6 -p_5\cdot p_6 \right]
\end{align}
For the last two diagrams we have used the $\phi$ quintic and sextic vertices \eqref{eq:phivertices}.
Summing over all 720 permutations of the external legs and combining the diagrams with the following symmetry factors
\begin{equation}\label{eq:treexxphi}
A^{\phi\phi\phi} \propto \frac18\, d_1^{\phi\phi\phi} + \frac{1}{16}\, d_2^{\phi\phi\phi} + \frac{1}{48}\, d_3 + \frac{1}{12}\, d_4^{\phi\phi\phi} + \frac{1}{72}\, d_5^{\phi\phi\phi} + \frac{1}{48}\, d_6^{\phi\phi\phi} + \frac{1}{720}\, d_7^{\phi\phi\phi} + \mathrm{perms} = 0
\end{equation}
it is straightforward to ascertain, e.g. numerically, that the amplitude vanishes for generic external momenta.

As before the only non-vanishing contribution comes from the kinematically singular configurations.
In particular the first, fourth and fifth diagrams contain a propagator which goes on-shell for forward kinematics. As for the gluons case one can group these three contributions and interpret them in terms of products of the diagrams in Figure \ref{fig:treephiphi}. In particular the first diagram receives contributions only from the first three diagrams of Figure \ref{fig:treephiphi}. The fourth diagram produces the products of the four-vertex interactions in the two-body amplitudes and the fifth diagram generates the mixed terms.
For instance, we can select the singular diagrams contributing to the structure $T_{12}T_{23}$. 
We dub $\hat d^{\phi\phi\phi}_i(\{p_j\})$ the diagrams listed above after removing the singular propagator and with the momenta ordered as in its argument and $\bar p_i = p_i$ the outgoing momenta after enforcing the $\delta$ function from the singular propagator. Then the total contribution with momentum $p_2$ flowing in the singular propagator is proportional to the combination
\begin{align}
T_{12}T_{23} \propto & \frac14\, \hat d^{\phi\phi\phi}_1\left( \raisebox{0.75mm}{$\{p_1, p_2$}, \raisebox{-0.75mm}{$\{p_3$}, \raisebox{0.75mm}{$\bar p_1\}$}, \raisebox{-0.75mm}{$\bar p_2, \bar p_3\}$} \right) + \frac{1}{12}\, \hat d^{\phi\phi\phi}_5\left( \raisebox{0.75mm}{$\{p_1, p_2$}, \raisebox{-0.75mm}{$\{p_3$}, \raisebox{0.75mm}{$\bar p_1\}$}, \raisebox{-0.75mm}{$\bar p_2, \bar p_3\}$} \right) + \nonumber\\& + \frac{1}{12}\, \hat d^{\phi\phi\phi}_5\left( \raisebox{0.75mm}{$\{\bar p_2, p_3$}, \raisebox{-0.75mm}{$\{p_1$}, \raisebox{0.75mm}{$\bar p_3\}$}, \raisebox{-0.75mm}{$p_2, \bar p_1\}$} \right) + \hat d^{\phi\phi\phi}_4\left( p_1, p_2, p_3, \bar p_1, \bar p_2, \bar p_3 \right)
\end{align}
where brackets stand for symmetrization and apply to separate groups of momenta in a self-explanatory notation. The symmetry factors take into account equivalent configurations.
Dividing by them as in the above formula we see that there is one contribution from diagram 4, corresponding to the product of the four-vertex diagrams of Figure \ref{fig:treephiphi} contributing to $T_{12}$ and $T_{23}$, respectively.
Diagram 1 produces 9 terms which emerge from the product of the three diagrams of Figure \ref{fig:treephiphi} with cubic vertices only.
Finally diagram 5 gives 6 terms from the mixed products. Altogether these combine to give the $4\times4 = 16$ terms from the product of two-body amplitudes.
Inserting the Jacobians from the momentum conservation $\delta$ functions and properly normalizing, we have ascertained that this combination gives precisely $T_{12}T_{23}$, as it can be obtained from formula \eqref{eq:treephiphi}. Summing the contributions to $T_{12}T_{13}$ and $T_{13}T_{23}$, altogether they combine to give the full factorization \eqref{eq:YB}.

\subsection{Scattering of two gluons and one meson}

Next we can consider the mixed process $x(p_1)x(p_2)\phi(p_3)\to x(p_4)x(p_5)\phi(p_6)$.
In this case there are 23 topologies of diagram contributing, shown schematically in Figure \ref{fig:treexxphi}, with possible permutations of the external legs.
\FIGURE{
\centering
\includegraphics[width=\textwidth]{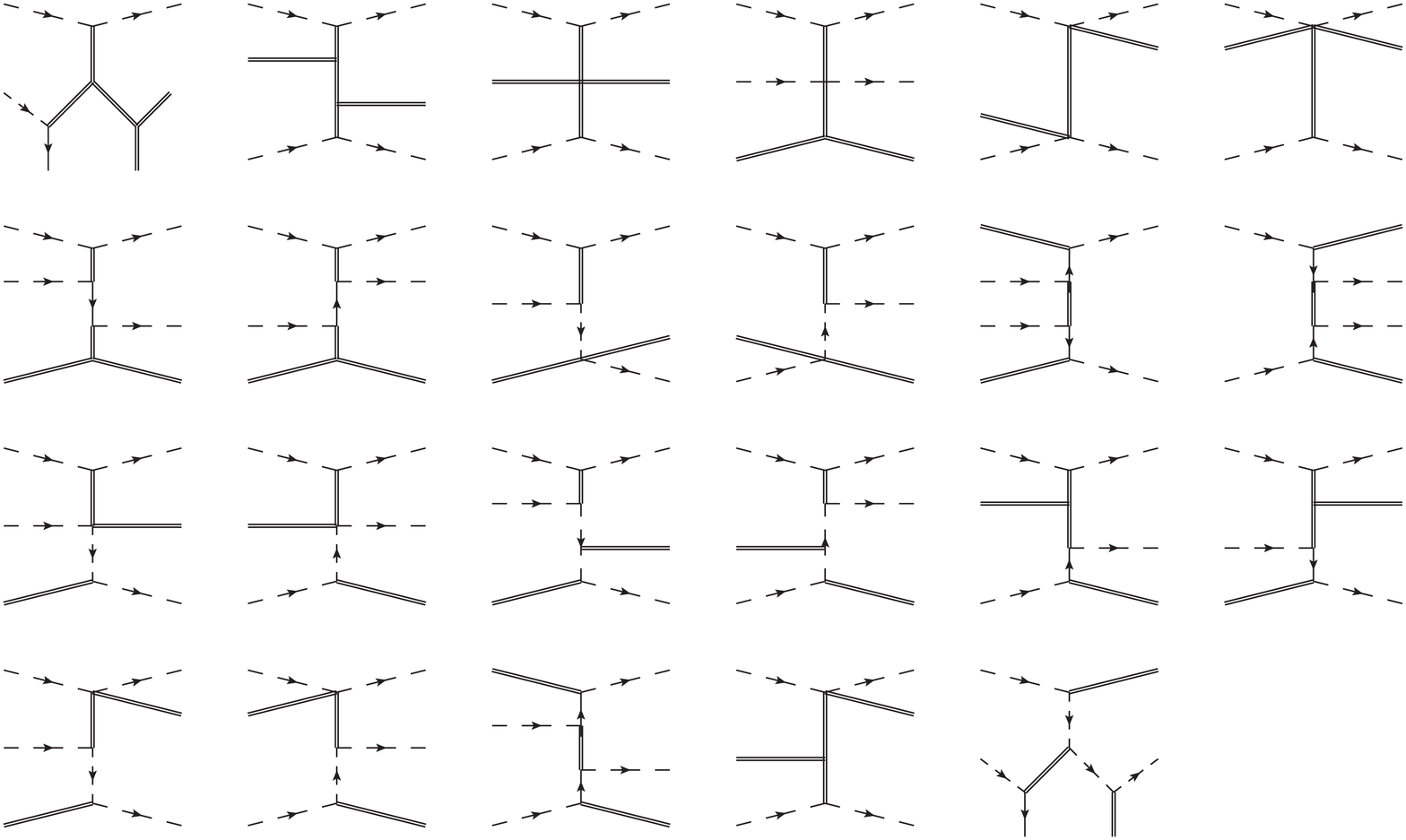}
\caption{Tree-level diagrams for $xx\phi \to xx\phi$ scattering.}
\label{fig:treexxphi}
}
In order to show factorization we evaluate this process numerically for generic configurations of external momenta satisfying the on-shell and momentum conservation conditions.
We use the expression for the diagrams in Appendix \ref{app:xxphi} and sum over the eight momentum permutations $p_1\leftrightarrow p_2$, $p_4\leftrightarrow p_5$ and $p_3\leftrightarrow p_6$.
For some diagrams these permutations are overcounting the contribution, which we take into account with the following symmetry factors
\begin{equation}
A^{xx\phi} \propto \frac14\, \sum_{i=1}^{2}\, d_i^{xx\phi} + \frac12\, \sum_{j=3}^{12}\, d_j^{xx\phi} + \sum_{k=13}^{23}\, d_k^{xx\phi} + \mathrm{perms} = 0
\end{equation}
Remarkably, such a large combination of diagrams can be straightforwardly seen to vanish for generic choices of external momenta, with a marvellous cancellations spreading over 132 terms.
Therefore only singular configurations corresponding to factorization of the amplitude eventually contribute.
In this case the amplitude factorises in the contributions $T^{xx}_{12}T^{x\phi}_{23}$, $T^{xx}_{12}T^{x\phi}_{13}$ and $T^{x\phi}_{13}T^{x\phi}_{23}$.
We have verified both diagrammatically and analytically that the first and second terms arise when combining diagrams 7, 8, 9, 10, 15 and 16 in the singular momentum configurations.
Finally we have ascertained that the last product of two-body amplitudes emerges from diagrams 2, 5, 11, 12, 17, 18, 19, 20, 21 and 22.

\subsection{Scattering of one gluon and two mesons}

Scattering of a gluon and two mesons receives contributions from 29 topologies of Feynman diagrams, depicted in Figure \ref{fig:treexphiphi}.
In each there are up to $4!$ factorial permutations of the external momenta of the mesons.
\FIGURE{
\centering
\includegraphics[width=\textwidth]{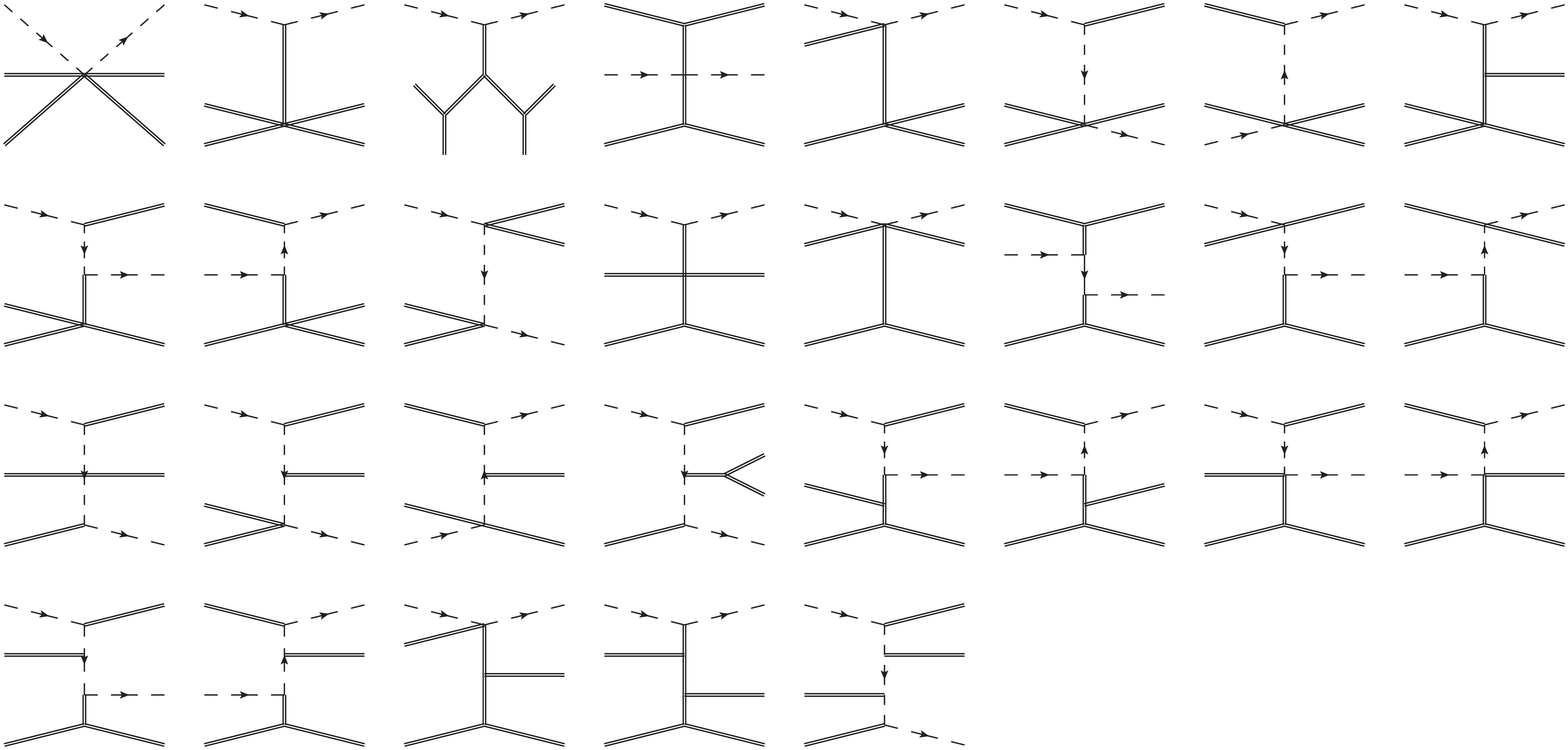}
\caption{Tree-level diagrams for $x\phi\phi \to x\phi\phi$ scattering.}
\label{fig:treexphiphi}
}
We take them into account by summing all diagrams over these permutations of momenta and dividing by the symmetry factors
\begin{equation}\label{eq:treexphiphi}
A^{x\phi\phi} \propto \frac{1}{4!}\, \sum_{i=1}^{2}\, d_i^{x\phi\phi} + \frac{1}{8}\, \sum_{i=3}^{4}\, d_i^{x\phi\phi} + \frac{1}{6}\, \sum_{i=5}^{10}\, d_i^{x\phi\phi} + \frac14\, \sum_{j=11}^{16}\, d_j^{x\phi\phi} + \frac12 \sum_{k=17}^{28}\, d_k^{x\phi\phi} + d_{29}^{x\phi\phi} + \mathrm{perms} = 0
\end{equation}
following the order in the figure. The contributions $d^{x\phi\phi}$ are collected in Appendix \ref{app:xphiphi}.
These are 236 contributions and we verified they sum to 0 for generic momenta configurations, providing a strong test of absence of particle production of the model.
The singular momentum configurations affecting diagrams 5, 8, 9, 10, 21, 22, 27 and 28 and 11, 14, 15, 16, 18, 19, 25, 26 and 29 combine to give the contributions $T^{x\phi}_{12}T^{\phi\phi}_{23}$ and $T^{x\phi}_{13}T^{\phi\phi}_{23}$, and $T^{x\phi}_{12}T^{x\phi}_{13}$, respectively. This proves that the amplitude factorises.

\section{Conclusions}

In this paper we have computed S-matrix elements for the excitations of the GKP string at first order in $1/g$ from perturbation theory of the light-cone gauge-fixed $AdS_5\times S^5$ sigma model.
The outcome of our analysis is that, as long as massless scalars do not enter the computation, the scattering phases are in agreement with the ABA predictions.
This safe sector includes all amplitudes without massless scalars on the external legs, apart from the fermion-fermion scattering process where massless scalar exchanges contribute.
In the latter case the result of a naive perturbative computation is found to violate the $SU(4)$ symmetry of the integrability based result, since its tensor structure does not consist of invariant tensors only. A possible interpretation of this fact is that IR singularities appearing at higher orders in the perturbative expansion spoil the predictivity of perturbation theory at tree level. Nevertheless, by comparing the perturbative results for the two $SU(4)$ invariant tensor structures and imposing that the spurious ones vanish, it is possible to correctly reproduce the scalar factor predicted by integrability. This hints at the fact that IR divergent contributions at higher loops should contribute only to the spurious tensor structures. It would be interesting to check this fact explicitly.

In an integrable theory $2\to 2$ processes are the fundamental building blocks for any higher point scattering amplitude thanks to the factorization of the S-matrix and the absence of particle production. We have explicitly checked these properties to hold for three-body S-matrices involving gluons and mesons. The structure of the computation turned out to be more involved than the BMN case \cite{Klose:2007rz}, where only quartic and sextic interactions are present. Here, also three- and five-point vertices need to be included and this considerably increases the number of diagrams. Therefore, the precise cancellation of the three-body S-matrix provides a further stringent check of the integrability of the model.

We conclude remarking that a similar analysis could be performed for the analogous $AdS_4\times \mathbb{CP}^3$ model dual to the ABJM theory.
Again it is expected that only a subset of these amplitudes is safely computable and comparable to the integrability predictions. In particular the latter model includes a massless Dirac fermion as well, whose dynamics is expected to be deeply nonperturbative, as for the massless scalars.
Finally, we point out that the tree level scattering elements we have computed (or the more comprehensive list from the ABA) could be used as the starting point of a unitarity based computation of the scattering phases at next order, in order to perform more precise checks of integrability of the S-matrix at the quantum level.
This program has been already applied to the BMN string in several $AdS$ backgrounds \cite{Engelund:2013fja,Bianchi:2013nra,Bianchi:2014rfa,Engelund:2014pla,Hoare:2014kma} and it would be interesting to extend it to the GKP string as well.

\section*{Acknowledgements}

We thank Benjamin Basso, Valentina Forini, Ben Hoare, Simone Piscaglia and Marco Rossi for very useful discussions. 
The work of LB is funded by DFG via the Emmy Noether Program
``Gauge Fields from Strings''. 
The work of MB was supported in part by the Science and Technology Facilities Council Consolidated Grant ST/L000415/1 \emph{String theory, gauge theory \& duality}.

\vfill
\newpage

\appendix

\section{Expanded Lagrangian to fourth order}
\label{app:lagr_exp}

In this appendix we spell out the interaction terms of the Lagriangin \eqref{eq:lagrangian}, up to quartic order in the fields.
Cubic vertices read
\begin{align}
{\cal L}_3 &=
-4\tilde\phi\, |\partial_s x - x|^2 + 2 \phi [(\partial_t \phi)^2-(\partial_s \phi)^2] + 2 \phi\ [(\partial_t y^a)^2-(\partial_s y^a)^2] + \nonumber\\
&
+ 4i\, \phi [(\partial_s \bar\psi_i - \bar\psi_i) \Pi_{+} \psi^i + \bar\psi_i \Pi_{-} (\partial_s \psi^i - \psi^i)] + \nonumber\\
&
+ 2 i\, y^a [(\partial_s \bar\psi_i - \bar\psi_i) \P_{+} (\rho^{a6})^{i}_{\phantom{i}j} \psi^j - \bar\psi_i \Pi_{-} (\rho^{a6})^{i}_{\phantom{i}j} (\partial_s \psi^j - \psi^j) ] + 2i\, \partial_t y^a \bar\psi_i \gamma^t \P_{+} (\rho^{a6})^{i}_{\phantom{i}j} \psi^j +  \nonumber\\&
+ 2 (\partial_s x - x) (\psi^i)^T \P_{+} (\rho^6)_{ij} \psi^j - 2 (\partial_s x^* - x^*) \bar\psi_i \P_{-} (\rho^{\dagger}_6)^{ij} (\bar\psi_j)^T
\end{align}
and quartic interactions
\begin{align}
{\cal L}_4 &=
8\, \phi^2\, |\partial_s  x - x|^2 + 2\, \phi^2
[\partial_{\alpha} \phi \partial_{\alpha} \phi + \frac{2}{3} \phi^2]
+ 2 \phi^2 \partial_{\alpha} y^a\partial_{\alpha} y^a - \frac{1}{2}y^a y^a\, \partial_{\alpha} y^b \partial_{\alpha} y^b + \nonumber\\
&
- i (4\phi^2 -y^a y^a)\,[(\partial_s \bar\psi_i - \bar\psi_i) \Pi_{+} \psi^i + \bar\psi_i \Pi_{-} (\partial_s \psi^i - \psi^i)] + \nonumber\\&
- 4 i\, \phi\,y^a [(\partial_s \bar\psi_i - \bar\psi_i) \P_{+} (\rho^{a6})^{i}_{\phantom{i}j} \psi^j - \bar\psi_i \Pi_{-} (\rho^{a6})^{i}_{\phantom{i}j} (\partial_s \psi^j - \psi^j)] + \nonumber\\
&
- 6 \phi\,[(\partial_s x - x) (\psi^i)^T \P_{+} (\rho^6)_{ij} \psi^j - (\partial_s x^* - x^*) \bar\psi_i \P_{-} (\rho^{\dagger}_6)^{ij} (\bar\psi_j)^T] + \nonumber\\
&
+ 2(\partial_s x - x) (\psi^i)^T \P_{+} (\rho^a)_{ij} y^a \psi^j - 2(\partial_s x^* - x^*) \bar\psi_i \P_{-} (\rho^{\dagger}_a)^{ij} y^a (\bar\psi_j)^T + \nonumber\\
&
- 2 i\, y^a\partial_t y^b\, \bar\psi_i \gamma^t \P_{+} (\rho^{ab})^{i}_{\phantom{i}j} \psi^j + (\bar\psi_i \gamma^t \P_{+} (\rho^{a6})^{i}_{\phantom{i}j} \psi^j)^2 - (\bar\psi_i \gamma^t \P_{+} \psi^i)^2
\end{align}
In the computation of scattering of three mesons quintic and sextic vertices are needed, which can be obtained expanding \eqref{eq:lagrangian}
\begin{equation}\label{eq:phivertices}
{\cal L}^{x,\phi}_{5,6} = -\frac{32}{3}\, \phi^3\, \big| \partial_s x - x \big|^2 + \frac{32}{3}\, \phi^4\, \big| \partial_s x - x \big|^2 + \frac43\, \left((\partial_t \phi)^2 - (\partial_s \phi)^2 \right) \phi^3 + \left(\frac{8}{45}\, \phi^2 + \frac23\, (\partial_\alpha \phi)^2\right)\phi^4
\end{equation}

\section{Computation of $xx\phi\to xx\phi$ diagrams}
\label{app:xxphi}

In this section we spell out the expressions of the diagrams contributing to the $xx\phi \to xx\phi$ scattering process.
We label external momenta as $x(p_1)x(p_2)\phi(p_3)\to x(p_4)x(p_5)\phi(-p_6)$ and give the expression for the diagrams of Figure \ref{fig:treexxphi}, following their order.
The diagrams have an overall factor $2g(i p_1-1)(i p_2-1)(-i p_4-1)(-i p_5-1)$ which we omit in the following. The remaining expressions read
\begin{align}
d^{xx\phi}_1 & = -\frac{16\left[ 4 - (p_{14}\cdot p_{25}+p_3\cdot p_{14}+p_3\cdot p_{14}+p_6\cdot p_{25}+p_6\cdot p_{25}+p_3\cdot p_6)\right]}{\left[(p_1-p_4)^2+4\right]\left[(p_2-p_5)^2+4\right]} \nonumber\\
d^{xx\phi}_2 & = \frac{16\left[ e_{14}e_{25} + e_{25}^2 + e_{14}^2  - (e\leftrightarrow \ppp)\right]\left[ e_3 (e_3+e_6) + e_3^2 + e_6^2  - (e\leftrightarrow \ppp)\right]}{\left[(p_1-p_4)^2+4\right]\left[(p_3+p_6)^2+4\right]\left[(p_2-p_5)^2+4\right]} \nonumber\\
d^{xx\phi}_3 & = \frac{16\left[ e_3 e_{14} + e_3^2 + e_{14}^2  - (e\leftrightarrow \ppp)\right]\left[ e_6 e_{25} + e_6^2 + e_{25}^2  - (e\leftrightarrow \ppp)\right]}{\left[(p_1-p_4)^2+4\right]\left[(p_1+p_3-p_4)^2+4\right]\left[(p_2-p_5)^2+4\right]} \nonumber\\
d^{xx\phi}_4 & = \frac{32\left[ e_3 e_6 + e_3^2 + e_6^2  - (e\leftrightarrow \ppp)\right]}{\left[(p_1-p_4)^2+4\right]\left[(p_3+p_6)^2+4\right]} \nonumber\\
d^{xx\phi}_5 & = \frac{64}{\left[(p_1+p_3-p_4)^2+4\right]} \nonumber\\
d^{xx\phi}_6 & = \frac{64}{\left[(p_2-p_5)^2+4\right]} \nonumber\\
d^{xx\phi}_7 & = \frac{32\left[ -e_3 e_6 - e_3^2 - e_6^2  - (e\leftrightarrow \ppp)\right]\left[(\ppp_1+\ppp_2-\ppp_4)^2+1\right]}{\left[(p_1-p_4)^2+4\right]\left[(p_3+p_6)^2+4\right]\left[(p_1+p_2-p_4)^2+2\right]} \nonumber\\
d^{xx\phi}_8 & = \frac{32\left[ -e_3 e_6 - e_3^2 - e_6^2  - (e\leftrightarrow \ppp)\right]\left[(\ppp_2+\ppp_3+\ppp_6)^2+1\right]}{\left[(p_1-p_4)^2+4\right]\left[(p_3+p_6)^2+4\right]\left[(p_2+p_3+p_6)^2+2\right]} \nonumber\\
d^{xx\phi}_9 & = -\frac{64\left[(\ppp_1+\ppp_2-\ppp_4)^2+1\right]}{\left[(p_1-p_4)^2+4\right]\left[(p_1+p_2-p_4)^2+2\right]} \nonumber\\
d^{xx\phi}_{10} & = -\frac{64\left[(\ppp_2+\ppp_3+\ppp_6)^2+1\right]}{\left[(p_1-p_4)^2+4\right]\left[(p_2+p_3+p_6)^2+2\right]} \nonumber\\
d^{xx\phi}_{11} & = \frac{64\left[(\ppp_4-\ppp_3)^2+1\right]\left[(\ppp_5-\ppp_6)^2+1\right]}{\left[(p_1+p_3-p_4)^2+4\right]\left[(p_4-p_3)^2+2\right]\left[(p_5-p_6)^2+2\right]} \nonumber\\
d^{xx\phi}_{12} & = \frac{64\left[(\ppp_1+\ppp_3)^2+1\right]\left[(\ppp_2+\ppp_6)^2+1\right]}{\left[(p_1+p_3-p_4)^2+4\right]\left[(p_1+p_3)^2+2\right]\left[(p_2+p_6)^2+2\right]} \nonumber\\
d^{xx\phi}_{13} & = -\frac{64\left[(\ppp_5-\ppp_6)^2+1\right]}{\left[(p_1-p_4)^2+4\right]\left[(p_5-p_6)^2+2\right]} \nonumber\\
d^{xx\phi}_{14} & = -\frac{64\left[(\ppp_2+\ppp_6)^2+1\right]}{\left[(p_1-p_4)^2+4\right]\left[(p_2+p_6)^2+2\right]} \nonumber\\
d^{xx\phi}_{15} & = \frac{64\left[(\ppp_5-\ppp_6)^2+1\right]\left[(\ppp_5-\ppp_3-\ppp_6)^2+1\right]}{\left[(p_1-p_4)^2+4\right]\left[(p_5-p_3-p_6)^2+2\right]\left[(p_5-p_6)^2+2\right]} \nonumber\\
d^{xx\phi}_{16} & = \frac{64\left[(\ppp_2+\ppp_6)^2+1\right]\left[(\ppp_2+\ppp_3+\ppp_6)^2+1\right]}{\left[(p_1-p_4)^2+4\right]\left[(p_2+p_3+p_6)^2+2\right]\left[(p_2+p_6)^2+2\right]} \nonumber\\
d^{xx\phi}_{17} & = -\frac{32\left[ (e_1-e_4) e_3 + (e_1-e_4)^2 + e_3^2 - (e\leftrightarrow \ppp)\right]\left[(\ppp_2+\ppp_6)^2+1\right]}{\left[(p_1-p_4)^2+4\right]\left[(p_1+p_3-p_4)^2+4\right]\left[(p_2+p_6)^2+2\right]} \nonumber\\
d^{xx\phi}_{18} & = -\frac{32\left[ (e_1-e_4) e_3 + (e_1-e_4)^2 + e_3^2 - (e\leftrightarrow \ppp)\right]\left[(\ppp_5-\ppp_6)^2+1\right]}{\left[(p_1-p_4)^2+4\right]\left[(p_1+p_3-p_4)^2+4\right]\left[(p_5-p_6)^2+2\right]} \nonumber\\
d^{xx\phi}_{19} & = -\frac{64\left[(\ppp_3-\ppp_5)^2+1\right]}{\left[(p_1+p_6-p_4)^2+4\right]\left[(p_5-p_3)^2+2\right]} \nonumber\\
d^{xx\phi}_{20} & = -\frac{64\left[(\ppp_2+\ppp_6)^2+1\right]}{\left[(p_1+p_3-p_4)^2+4\right]\left[(p_2+p_6)^2+2\right]} \nonumber\\
d^{xx\phi}_{21} & = \frac{64\left[(\ppp_4-\ppp_3)^2+1\right]\left[(\ppp_2+\ppp_6)^2+1\right]}{\left[(p_1+p_3-p_4)^2+4\right]\left[(p_4-p_3)^2+2\right]\left[(p_2+p_6)^2+2\right]} \nonumber\\
d^{xx\phi}_{22} & = \frac{32\left[ -(e_1-e_4+e_3) (e_2-e_5) - (e_1-e_4+e_3) e_6 - (e_2-e_5) e_6 - (e\leftrightarrow \ppp)\right]}{\left[(p_1+p_3-p_4)^2+4\right]\left[(p_2-p_5)^2+4\right]} \nonumber\\
d^{xx\phi}_{23} & = \frac{64\left[(\ppp_1+\ppp_6)^2+1\right]\left[(\ppp_3-\ppp_4)^2+1\right]}{\left[(p_2-p_5)^2+4\right]\left[(p_1+p_6)^2+2\right]\left[(p_3-p_4)^2+2\right]} \nonumber\\
\end{align}

\section{Computation of $x\phi\phi\to x\phi\phi$ diagrams}
\label{app:xphiphi}

In this section we give the expressions for the contributions $d^{x\phi\phi}$ relevant for $x\phi\phi\to x\phi\phi$ scattering.
The corresponding diagrams are shown in Figure \ref{fig:treexphiphi}.
The incoming $x$ particle has momentum $p_1$ and the outgoing $p_4$.
An overall common factor $2g(i \ppp_1-1)(-i \ppp_4-1)$ is understood in the following formulae.
The $\phi$ particles have momenta $p_2$, $p_3$, $p_5$ and $p_6$ and we take them as all ingoing for simplicity.
The following formulae hold for a sample configuration of momenta for the mesons and have to be symmetrised in the corresponding momentum indices.
These are 24 permutations which, for all but the last contribution, overcount the diagram by a symmetry factor which we divide by in \eqref{eq:treexphiphi}.
\begin{align}
d^{x\phi\phi}_1 &= -64  \nonumber\\
d^{x\phi\phi}_2 &= \frac{16\left[e_{14} e_2 + e_{14} e_3 + e_{14} e_5 + e_{14} e_6 + e_2 e_3 + e_2 e_5 + e_2 e_6 + e_3 e_5 + e_3 e_6 + e_5 e_6
- (e\leftrightarrow \ppp)\right]}{(p_1-p_4)^2 + 4}
\nonumber\\
d^{x\phi\phi}_3 &= -\frac{16 \left[ - e_{14}(e_2+e_5) - e_{14}(e_3+e_6) - (e_2+e_5) (e_3+e_6) - (e\leftrightarrow \ppp)\right]}{\left[ p_{14}^2 + 4 \right]\left[ (p_2+p_5)^2 + 4 \right] \left[ (p_3+p_6)^2 + 4 \right]}\nonumber\\&\left[ e_2^2 + e_5^2 + e_2 e_5 - (e\leftrightarrow \ppp)\right]
\left[e_3^2 + e_6^2 + e_3 e_6 - (e\leftrightarrow \ppp)\right]  \nonumber\\
d^{x\phi\phi}_4 &= -\frac{32\left[ e_2^2 + e_5^2 + e_2 e_5 - (e\leftrightarrow \ppp)\right]}{\left[ (p_2+p_5)^2 + 4 \right]\left[ (p_3+p_6)^2 + 4 \right]}
\left[  e_3^2 + e_6^2 + e_3 e_6 - (e\leftrightarrow \ppp) \right]  \nonumber\\
d^{x\phi\phi}_5 &= \frac{32}{\left[ (p_1+p_2-p_4)^2 + 4 \right]}\left[ 4 - ((p_1+p_2-p_4)\cdot p_3 + (p_1+p_2-p_4)\cdot p_5 + \right.\nonumber\\&\left. + (p_1+p_2-p_4)\cdot p_6 + p_3\cdot p_5 + p_3\cdot p_6 + p_5\cdot p_6 )\right] \nonumber\\
d^{x\phi\phi}_6 &= \frac{128\left[ (\ppp_1+\ppp_2)^2 + 1 \right]}{\left[ (p_1+p_2)^2 + 2 \right]} \nonumber\\
d^{x\phi\phi}_7 &= \frac{128\left[ (\ppp_4-\ppp_2)^2 + 1 \right]}{\left[ (p_4-p_2)^2 + 2 \right]} \nonumber\\
d^{x\phi\phi}_8 &= \frac{16\left[
e_{14} e_2 + e_{14} (e_3+e_5+e_6) + e_2 (e_3+e_5+e_6) - (e\leftrightarrow \ppp)\right]}{\left[ (p_1-p_4)^2 + 4 \right]\left[ (p_3+p_5+p_6)^2 + 4 \right]}
\left[ 4 - ( p_3\cdot p_6 + p_5\cdot p_6 + \right.\nonumber\\&\left. -(p_3+p_5+p_6)\cdot p_3 -(p_3+p_5+p_6)\cdot p_5 -(p_3+p_5+p_6)\cdot p_6 + p_3\cdot p_5 )\right] \nonumber\\
d^{x\phi\phi}_9 &= -\frac{32\left[ (\ppp_1+\ppp_2)^2 + 1 \right]}{\left[ (p_1+p_2)^2 + 2 \right]\left[ (p_3+p_5+p_6)^2 + 4 \right]}\left[ 4 - (-(p_3+p_5+p_6)\cdot p_3 + \right.\nonumber\\&\left. -(p_3+p_5+p_6)\cdot p_5 -(p_3+p_5+p_6)\cdot p_6 + p_3\cdot p_5 + p_3\cdot p_6 + p_5\cdot p_6 )\right] \nonumber\\
d^{x\phi\phi}_{10} &= -\frac{32\left[ (\ppp_2-\ppp_4)^2 + 1 \right]}{\left[ (p_2-p_4)^2 + 2 \right]\left[ (p_3+p_5+p_6)^2 + 4 \right]}\left[ 4 - (-(p_3+p_5+p_6)\cdot p_3 + \right.\nonumber\\&\left. -(p_3+p_5+p_6)\cdot p_5 -(p_3+p_5+p_6)\cdot p_6 + p_3\cdot p_5 + p_3\cdot p_6 + p_5\cdot p_6 )\right] \nonumber\\
d^{x\phi\phi}_{11} &= \frac{128\left[ (\ppp_1+\ppp_2+\ppp_3)^2 + 1 \right]}{\left[ (p_1+p_2+p_3)^2 + 2 \right]} \nonumber\\
d^{x\phi\phi}_{12} &= \frac{16\left[
e_3^2 + e_6^2 + e_3 e_6 - (e\leftrightarrow \ppp)\right] }{\left[ (p_1-p_4)^2 + 4 \right]\left[ (p_3+p_6)^2 + 4 \right]} \left[ 4 - ((p_1-p_4)\cdot p_2 + (p_1-p_4)\cdot (p_3+p_6) + \right.\nonumber\\&\left. + (p_1-p_4)\cdot p_5 + (p_3+p_6)\cdot p_2 + (p_3+p_6)\cdot p_5 + p_2\cdot p_5 )\right] \nonumber\\
d^{x\phi\phi}_{13} &= -\frac{64\left[
e_3^2 + e_6^2 + e_3 e_6 - (e\leftrightarrow \ppp)\right] }{\left[ (p_3+p_6)^2 + 4 \right]}  \nonumber\\
d^{x\phi\phi}_{14} &= \frac{32\left[ (\ppp_1+\ppp_2+\ppp_5)^2 + 1 \right]\left[ e_2^2 + e_5^2 + e_2 e_5 - (e\leftrightarrow \ppp)\right]}{\left[ (p_1+p_2+p_5)^2 + 2 \right]\left[ (p_2+p_5)^2 + 4 \right] \left[ (p_3+p_6)^2 + 4 \right]}\left[e_3^2 + e_6^2 + e_3 e_6 - (e\leftrightarrow \ppp)\right]\nonumber\\
d^{x\phi\phi}_{15} &= \frac{64\left[ (\ppp_1+\ppp_2+\ppp_5)^2 + 1 \right]\left[e_3^2 + e_6^2 + e_3 e_6 - (e\leftrightarrow \ppp)\right]}{\left[ (p_1+p_2+p_5)^2 + 2 \right]\left[ (p_3+p_6)^2 + 4 \right]}\nonumber\\
d^{x\phi\phi}_{16} &= \frac{64\left[ (\ppp_4-\ppp_2-\ppp_5)^2 + 1 \right]\left[e_3^2 + e_6^2 + e_3 e_6 - (e\leftrightarrow \ppp)\right]}{\left[ (p_4-p_2-p_5)^2 + 2 \right]\left[ (p_3+p_6)^2 + 4 \right]}\nonumber\\
d^{x\phi\phi}_{17} &= -\frac{128\left[ (\ppp_1+\ppp_2)^2 + 1 \right]\left[ (\ppp_1+\ppp_2+\ppp_3)^2 + 1 \right]}{\left[ (p_1+p_2)^2 + 2 \right]\left[ (p_1+p_2+p_3)^2 + 2 \right]}\nonumber\\
d^{x\phi\phi}_{18} &= -\frac{128\left[ (\ppp_4-\ppp_6)^2 + 1 \right]\left[ (\ppp_1+\ppp_2+\ppp_3)^2 + 1 \right]}{\left[ (p_4-p_6)^2 + 2 \right]\left[ (p_1+p_2+p_3)^2 + 2 \right]}\nonumber\\
d^{x\phi\phi}_{19} &= -\frac{64\left[ (\ppp_1+\ppp_2)^2 + 1 \right]\left[ (\ppp_4-\ppp_5)^2 + 1 \right]\left[e_3^2 + e_6^2 + e_3 e_6 - (e\leftrightarrow \ppp)\right]}{\left[ (p_1+p_2)^2 + 2 \right]\left[ (p_4-p_5)^2 + 2 \right]\left[ (p_3+p_6)^2 + 4 \right]}\nonumber\\
d^{x\phi\phi}_{20} &= -\frac{32\left[ (\ppp_1+\ppp_2)^2 + 1 \right]
\left[e_3^2 + e_6^2 + e_3 e_6 - (e\leftrightarrow \ppp)\right]}{\left[ (p_1+p_2)^2 + 2 \right]\left[ (p_1+p_2-p_4)^2 + 4 \right]\left[ (p_3+p_6)^2 + 4 \right]}\nonumber\\&
\left[(e_1+e_2-e_4) e_5 + (e_1+e_2-e_4)(e_3+e_6) + (e_3+e_6) e_5 - (e\leftrightarrow \ppp)\right]
\nonumber\\
d^{x\phi\phi}_{21} &= \frac{32\left[ (\ppp_4-\ppp_2)^2 + 1 \right]
\left[e_3^2 + e_6^2 + e_3 e_6 - (e\leftrightarrow \ppp)\right]}{\left[ (p_4-p_2)^2 + 2 \right]\left[ (p_3+p_5+p_6)^2 + 4 \right]\left[ (p_3+p_6)^2 + 4 \right]}\nonumber\\&
\left[(e_3+e_5+e_6) e_5 + (e_3+e_5+e_6)(e_3+e_6) - (e_3+e_6) e_5 - (e\leftrightarrow \ppp)\right]
\nonumber\\
d^{x\phi\phi}_{22} &= -\frac{16
\left[e_3^2 + e_6^2 + e_3 e_6 - (e\leftrightarrow \ppp)\right]}{\left[ (p_1-p_4)^2 + 4 \right]\left[ (p_1-p_4+p_2)^2 + 4 \right]\left[ (p_3+p_6)^2 + 4 \right]}\nonumber\\&
\left[- (e_1+e_2-e_4) (e_1-e_4) - (e_1+e_2-e_4) e_2 + (e_1-e_4) e_2 - (e\leftrightarrow \ppp)\right]\nonumber\\&\left[(e_1+p_2-p_4) (e_3+e_6) + (e_1+p_2-p_4) e_5 + (e_3+e_6) e_5 - (e\leftrightarrow \ppp)\right]
\nonumber\\
d^{x\phi\phi}_{23} &= \frac{32
\left[e_3^2 + e_6^2 + e_3 e_6 - (e\leftrightarrow \ppp)\right]}{\left[ (p_1-p_4+p_2)^2 + 4 \right]\left[ (p_3+p_6)^2 + 4 \right]}\nonumber\\&
\left[(e_1+e_2-e_4) (e_3+e_6) + (e_1+e_2-e_4) e_5 + (e_3+e_6) e_5 - (e\leftrightarrow \ppp)\right]
\nonumber\\
d^{x\phi\phi}_{24} &= -\frac{128
\left[ (\ppp_1+\ppp_2)^2 + 1 \right]\left[ (\ppp_4-\ppp_6)^2 + 1 \right]}{\left[ (p_1+p_2)^2 + 2 \right]\left[ (p_4-p_6)^2 + 2 \right]}
\nonumber\\
d^{x\phi\phi}_{25} &= -\frac{64
\left[ (\ppp_1+\ppp_2)^2 + 1 \right]\left[ (\ppp_1+\ppp_2+\ppp_5)^2 + 1 \right]}{\left[ (p_1+p_2)^2 + 2 \right]\left[ (p_1+p_2+p_5)^2 + 2 \right]\left[ (p_3+p_6)^2 + 4 \right]}\left[e_3^2 + e_6^2 + e_3 e_6 - (e\leftrightarrow \ppp)\right]
\nonumber\\
d^{x\phi\phi}_{26} &= -\frac{64
\left[ (\ppp_4-\ppp_2)^2 + 1 \right]\left[ (\ppp_4-\ppp_2-\ppp_5)^2 + 1 \right]}{\left[ (p_4-p_2)^2 + 2 \right]\left[ (p_4-p_2-p_5)^2 + 2 \right]\left[ (p_3+p_6)^2 + 4 \right]}\left[e_3^2 + e_6^2 + e_3 e_6 - (e\leftrightarrow \ppp)\right]
\nonumber\\
d^{x\phi\phi}_{27} &= \frac{64
\left[ (\ppp_1+\ppp_2)^2 + 1 \right]}{\left[ (p_1+p_2)^2 + 2 \right]\left[ (p_3+p_6)^2 + 4 \right]}\left[e_3^2 + e_6^2 + e_3 e_6 - (e\leftrightarrow \ppp)\right]
\nonumber\\
d^{x\phi\phi}_{28} &= \frac{64
\left[ (\ppp_4-\ppp_2)^2 + 1 \right]}{\left[ (p_4-p_2)^2 + 2 \right]\left[ (p_3+p_6)^2 + 4 \right]}\left[e_3^2 + e_6^2 + e_3 e_6 - (e\leftrightarrow \ppp)\right]
\nonumber\\
d^{x\phi\phi}_{29} &= \frac{128
\left[ (\ppp_1+\ppp_2)^2 + 1 \right]\left[ (\ppp_1+\ppp_2+\ppp_3)^2 + 1 \right]\left[ (\ppp_4-\ppp_5)^2 + 1 \right]}{\left[ (p_1+p_2)^2 + 2 \right]\left[ (p_1+p_2+p_3)^2 + 2 \right]\left[ (p_4-p_5)^2 + 2 \right]}
\end{align}

\bibliographystyle{JHEP}

\bibliography{biblio}

\providecommand{\href}[2]{#2}\begingroup\raggedright\begin{thebibliography}{10}

\bibitem{Maldacena:1997re}
J.~M. Maldacena, {\it {The Large N limit of superconformal field theories and
  supergravity}},  {\em Adv.Theor.Math.Phys.} {\bf 2} (1998) 231--252,
  [\href{http://xxx.lanl.gov/abs/hep-th/9711200}{{\tt hep-th/9711200}}].

\bibitem{Minahan:2002ve}
J.~Minahan and K.~Zarembo, {\it {The Bethe ansatz for N=4 superYang-Mills}},
  {\em JHEP} {\bf 0303} (2003) 013,
  [\href{http://xxx.lanl.gov/abs/hep-th/0212208}{{\tt hep-th/0212208}}].

\bibitem{Beisert:2003yb}
N.~Beisert and M.~Staudacher, {\it {The N=4 SYM integrable super spin chain}},
  {\em Nucl.Phys.} {\bf B670} (2003) 439--463,
  [\href{http://xxx.lanl.gov/abs/hep-th/0307042}{{\tt hep-th/0307042}}].

\bibitem{Beisert:2005fw}
N.~Beisert and M.~Staudacher, {\it {Long-range psu(2,2|4) Bethe Ansatze for
  gauge theory and strings}},  {\em Nucl.Phys.} {\bf B727} (2005) 1--62,
  [\href{http://xxx.lanl.gov/abs/hep-th/0504190}{{\tt hep-th/0504190}}].

\bibitem{Beisert:2006ez}
N.~Beisert, B.~Eden, and M.~Staudacher, {\it {Transcendentality and Crossing}},
   {\em J.Stat.Mech.} {\bf 0701} (2007) P01021,
  [\href{http://xxx.lanl.gov/abs/hep-th/0610251}{{\tt hep-th/0610251}}].

\bibitem{Gubser:2002tv}
S.~S. Gubser, I.~R. Klebanov, and A.~M. Polyakov, {\it A semi-classical limit
  of the gauge/string correspondence},  {\em Nucl. Phys.} {\bf B636} (2002)
  99--114, [\href{http://xxx.lanl.gov/abs/hep-th/0204051}{{\tt
  hep-th/0204051}}].

\bibitem{Frolov:2002av}
S.~Frolov and A.~A. Tseytlin, {\it Semiclassical quantization of rotating
  superstring in {$AdS_5 \times S^5$}},  {\em JHEP} {\bf 0206} (2002) 007,
  [\href{http://xxx.lanl.gov/abs/hep-th/0204226}{{\tt hep-th/0204226}}].

\bibitem{Metsaev:2000yu}
R.~Metsaev, C.~B. Thorn, and A.~A. Tseytlin, {\it {Light cone superstring in
  AdS space-time}},  {\em Nucl.Phys.} {\bf B596} (2001) 151--184,
  [\href{http://xxx.lanl.gov/abs/hep-th/0009171}{{\tt hep-th/0009171}}].

\bibitem{Metsaev:2000yf}
R.~Metsaev and A.~A. Tseytlin, {\it {Superstring action in AdS(5) x S**5. Kappa
  symmetry light cone gauge}},  {\em Phys.Rev.} {\bf D63} (2001) 046002,
  [\href{http://xxx.lanl.gov/abs/hep-th/0007036}{{\tt hep-th/0007036}}].

\bibitem{Giombi:2009gd}
S.~Giombi, R.~Ricci, R.~Roiban, A.~Tseytlin, and C.~Vergu, {\it {Quantum
  AdS(5)xS(5) superstring in the AdS light-cone gauge}},  {\em JHEP} {\bf 1003}
  (2010) 003, [\href{http://xxx.lanl.gov/abs/0912.5105}{{\tt
  arXiv:0912.5105}}].

\bibitem{Alday:2007mf}
L.~F. Alday and J.~M. Maldacena, {\it {Comments on operators with large spin}},
   {\em JHEP} {\bf 0711} (2007) 019,
  [\href{http://xxx.lanl.gov/abs/0708.0672}{{\tt arXiv:0708.0672}}].

\bibitem{Basso:2007wd}
B.~Basso, G.~P. Korchemsky, and J.~Kotanski, {\it {Cusp anomalous dimension in
  maximally supersymmetric Yang-Mills theory at strong coupling}},  {\em Phys.
  Rev. Lett.} {\bf 100} (2008) 091601,
  [\href{http://xxx.lanl.gov/abs/0708.3933}{{\tt arXiv:0708.3933}}].

\bibitem{Roiban:2007jf}
R.~Roiban, A.~Tirziu, and A.~A. Tseytlin, {\it {Two-loop world-sheet
  corrections in AdS(5)x S**5 superstring}},  {\em JHEP} {\bf 0707} (2007) 056,
  [\href{http://xxx.lanl.gov/abs/0704.3638}{{\tt arXiv:0704.3638}}].

\bibitem{Roiban:2007dq}
R.~Roiban and A.~A. Tseytlin, {\it {Strong-coupling expansion of cusp anomaly
  from quantum superstring}},  {\em JHEP} {\bf 0711} (2007) 016,
  [\href{http://xxx.lanl.gov/abs/0709.0681}{{\tt arXiv:0709.0681}}].

\bibitem{Roiban:2007ju}
R.~Roiban and A.~A. Tseytlin, {\it {Spinning superstrings at two loops:
  Strong-coupling corrections to dimensions of large-twist SYM operators}},
  {\em Phys.Rev.} {\bf D77} (2008) 066006,
  [\href{http://xxx.lanl.gov/abs/0712.2479}{{\tt arXiv:0712.2479}}].

\bibitem{Zarembo:2011ag}
K.~Zarembo and S.~Zieme, {\it {Fine Structure of String Spectrum in $AdS_5$ x
  $S^5$}},  {\em JETP Lett.} {\bf 95} (2012), no.~8 219--223,
  [\href{http://xxx.lanl.gov/abs/1110.6146}{{\tt arXiv:1110.6146}}].

\bibitem{Gromov:2008qe}
N.~Gromov and P.~Vieira, {\it {The all loop AdS4/CFT3 Bethe ansatz}},  {\em
  JHEP} {\bf 0901} (2009) 016, [\href{http://xxx.lanl.gov/abs/0807.0777}{{\tt
  arXiv:0807.0777}}].

\bibitem{Basso:2013pxa}
B.~Basso and A.~Rej, {\it {Bethe ansatze for GKP strings}},  {\em Nucl.Phys.}
  {\bf B879} (2014) 162--215, [\href{http://xxx.lanl.gov/abs/1306.1741}{{\tt
  arXiv:1306.1741}}].

\bibitem{Aharony:2008ug}
O.~Aharony, O.~Bergman, D.~L. Jafferis, and J.~Maldacena, {\it {$\mathcal{N} =
  6$} superconformal {C}hern-{S}imons-matter theories, {M2}-branes and their
  gravity duals},  {\em JHEP} {\bf 0810} (2008) 091,
  [\href{http://xxx.lanl.gov/abs/0806.1218}{{\tt arXiv:0806.1218}}].

\bibitem{Uvarov:2009hf}
D.~Uvarov, {\it {AdS(4) x CP**3 superstring in the light-cone gauge}},  {\em
  Nucl.Phys.} {\bf B826} (2010) 294--312,
  [\href{http://xxx.lanl.gov/abs/0906.4699}{{\tt arXiv:0906.4699}}].

\bibitem{Uvarov_main}
D.~Uvarov, {\it {Light-cone gauge Hamiltonian for AdS(4) x CP**3 superstring}},
   {\em Mod.Phys.Lett.} {\bf A25} (2010) 1251--1265,
  [\href{http://xxx.lanl.gov/abs/0912.1044}{{\tt arXiv:0912.1044}}].

\bibitem{Uvarov:2011zz}
D.~Uvarov, {\it {Light-cone gauge formulation for AdS(4) x CP(3) superstring}},
   {\em Phys.Part.Nucl.Lett.} {\bf 8} (2011) 272--278.

\bibitem{Bianchi:2014ada}
L.~Bianchi, M.~S. Bianchi, A.~Bres, V.~Forini, and E.~Vescovi, {\it {Two-loop
  cusp anomaly in ABJM at strong coupling}},  {\em JHEP} {\bf 1410} (2014) 13,
  [\href{http://xxx.lanl.gov/abs/1407.4788}{{\tt arXiv:1407.4788}}].

\bibitem{Bianchi:2015laa}
L.~Bianchi and M.~S. Bianchi, {\it {Quantum dispersion relations for the
  $AdS\_4 \times CP^3$ GKP string}},
  \href{http://xxx.lanl.gov/abs/1505.0078}{{\tt arXiv:1505.0078}}.

\bibitem{Fioravanti:2013eia}
D.~Fioravanti, S.~Piscaglia, and M.~Rossi, {\it {On the scattering over the GKP
  vacuum}},  {\em Phys.Lett.} {\bf B728} (2014) 288--295,
  [\href{http://xxx.lanl.gov/abs/1306.2292}{{\tt arXiv:1306.2292}}].

\bibitem{Fioravanti:2015dma}
D.~Fioravanti, S.~Piscaglia, and M.~Rossi, {\it {Asymptotic Bethe Ansatz on the
  GKP vacuum as a defect spin chain: scattering, particles and minimal area
  Wilson loops}},  {\em Nucl. Phys.} {\bf B898} (2015) 301--400,
  [\href{http://xxx.lanl.gov/abs/1503.0879}{{\tt arXiv:1503.0879}}].

\bibitem{Alday:2010ku}
L.~F. Alday, D.~Gaiotto, J.~Maldacena, A.~Sever, and P.~Vieira, {\it {An
  Operator Product Expansion for Polygonal null Wilson Loops}},  {\em JHEP}
  {\bf 1104} (2011) 088, [\href{http://xxx.lanl.gov/abs/1006.2788}{{\tt
  arXiv:1006.2788}}].

\bibitem{Gaiotto:2011dt}
D.~Gaiotto, J.~Maldacena, A.~Sever, and P.~Vieira, {\it {Pulling the straps of
  polygons}},  {\em JHEP} {\bf 1112} (2011) 011,
  [\href{http://xxx.lanl.gov/abs/1102.0062}{{\tt arXiv:1102.0062}}].

\bibitem{Basso:2013vsa}
B.~Basso, A.~Sever, and P.~Vieira, {\it {Spacetime and Flux Tube S-Matrices at
  Finite Coupling for N=4 Supersymmetric Yang-Mills Theory}},  {\em
  Phys.Rev.Lett.} {\bf 111} (2013), no.~9 091602,
  [\href{http://xxx.lanl.gov/abs/1303.1396}{{\tt arXiv:1303.1396}}].

\bibitem{Alday:2007hr}
L.~F. Alday and J.~M. Maldacena, {\it {Gluon scattering amplitudes at strong
  coupling}},  {\em JHEP} {\bf 06} (2007) 064,
  [\href{http://xxx.lanl.gov/abs/0705.0303}{{\tt arXiv:0705.0303}}].

\bibitem{Drummond:2007au}
J.~M. Drummond, J.~Henn, G.~P. Korchemsky, and E.~Sokatchev, {\it {Conformal
  Ward identities for Wilson loops and a test of the duality with gluon
  amplitudes}},  {\em Nucl. Phys.} {\bf B826} (2010) 337--364,
  [\href{http://xxx.lanl.gov/abs/0712.1223}{{\tt arXiv:0712.1223}}].

\bibitem{Drummond:2007cf}
J.~M. Drummond, J.~Henn, G.~P. Korchemsky, and E.~Sokatchev, {\it {On planar
  gluon amplitudes/Wilson loops duality}},  {\em Nucl. Phys.} {\bf B795} (2008)
  52--68, [\href{http://xxx.lanl.gov/abs/0709.2368}{{\tt arXiv:0709.2368}}].

\bibitem{Brandhuber:2007yx}
A.~Brandhuber, P.~Heslop, and G.~Travaglini, {\it {MHV amplitudes in N=4 super
  Yang-Mills and Wilson loops}},  {\em Nucl. Phys.} {\bf B794} (2008) 231--243,
  [\href{http://xxx.lanl.gov/abs/0707.1153}{{\tt arXiv:0707.1153}}].

\bibitem{Basso:2013aha}
B.~Basso, A.~Sever, and P.~Vieira, {\it {Space-time S-matrix and Flux tube
  S-matrix II. Extracting and Matching Data}},  {\em JHEP} {\bf 1401} (2014)
  008, [\href{http://xxx.lanl.gov/abs/1306.2058}{{\tt arXiv:1306.2058}}].

\bibitem{Basso:2014koa}
B.~Basso, A.~Sever, and P.~Vieira, {\it {Space-time S-matrix and Flux-tube
  S-matrix III. The two-particle contributions}},  {\em JHEP} {\bf 1408} (2014)
  085, [\href{http://xxx.lanl.gov/abs/1402.3307}{{\tt arXiv:1402.3307}}].

\bibitem{Basso:2014jfa}
B.~Basso, A.~Sever, and P.~Vieira, {\it {Collinear Limit of Scattering
  Amplitudes at Strong Coupling}},  {\em Phys.Rev.Lett.} {\bf 113} (2014),
  no.~26 261604, [\href{http://xxx.lanl.gov/abs/1405.6350}{{\tt
  arXiv:1405.6350}}].

\bibitem{Basso:2014nra}
B.~Basso, A.~Sever, and P.~Vieira, {\it {Space-time S-matrix and Flux-tube
  S-matrix IV. Gluons and Fusion}},  {\em JHEP} {\bf 1409} (2014) 149,
  [\href{http://xxx.lanl.gov/abs/1407.1736}{{\tt arXiv:1407.1736}}].

\bibitem{Basso:2014hfa}
B.~Basso, J.~Caetano, L.~Cordova, A.~Sever, and P.~Vieira, {\it {OPE for all
  Helicity Amplitudes}},  \href{http://xxx.lanl.gov/abs/1412.1132}{{\tt
  arXiv:1412.1132}}.

\bibitem{Basso:2015rta}
B.~Basso, J.~Caetano, L.~Cordova, A.~Sever, and P.~Vieira, {\it {OPE for all
  Helicity Amplitudes II. Form Factors and Data analysis}},
  \href{http://xxx.lanl.gov/abs/1508.0298}{{\tt arXiv:1508.0298}}.

\bibitem{Basso:2015uxa}
B.~Basso, A.~Sever, and P.~Vieira, {\it {Hexagonal Wilson Loops in Planar
  $\mathcal{N}=4$ SYM Theory at Finite Coupling}},
  \href{http://xxx.lanl.gov/abs/1508.0304}{{\tt arXiv:1508.0304}}.

\bibitem{Giombi:2010bj}
S.~Giombi, R.~Ricci, R.~Roiban, and A.~Tseytlin, {\it Quantum dispersion
  relations for excitations of long folded spinning superstring in {$AdS_5
  \times S^5$}},  {\em JHEP} {\bf 1101} (2011) 128,
  [\href{http://xxx.lanl.gov/abs/1011.2755}{{\tt arXiv:1011.2755}}].

\bibitem{Klose:2007rz}
T.~Klose, T.~McLoughlin, J.~Minahan, and K.~Zarembo, {\it {World-sheet
  scattering in AdS(5) x S**5 at two loops}},  {\em JHEP} {\bf 0708} (2007)
  051, [\href{http://xxx.lanl.gov/abs/0704.3891}{{\tt arXiv:0704.3891}}].

\bibitem{Zamolodchikov:2013ama}
A.~Zamolodchikov, {\it {Ising Spectroscopy II: Particles and poles at $T >
  T_c$}},  \href{http://xxx.lanl.gov/abs/1310.4821}{{\tt arXiv:1310.4821}}.

\bibitem{Berg:1977dp}
B.~Berg, M.~Karowski, P.~Weisz, and V.~Kurak, {\it {Factorized U(n) Symmetric s
  Matrices in Two-Dimensions}},  {\em Nucl. Phys.} {\bf B134} (1978) 125.

\bibitem{Zamolodchikov:1978xm}
A.~B. Zamolodchikov and A.~B. Zamolodchikov, {\it {Factorized s Matrices in
  Two-Dimensions as the Exact Solutions of Certain Relativistic Quantum Field
  Models}},  {\em Annals Phys.} {\bf 120} (1979) 253--291.

\bibitem{Engelund:2013fja}
O.~T. Engelund, R.~W. McKeown, and R.~Roiban, {\it {Generalized unitarity and
  the worldsheet $S$ matrix in $AdS_n \times S^n \times M^{10-2n}$}},  {\em
  JHEP} {\bf 1308} (2013) 023, [\href{http://xxx.lanl.gov/abs/1304.4281}{{\tt
  arXiv:1304.4281}}].

\bibitem{Bianchi:2013nra}
L.~Bianchi, V.~Forini, and B.~Hoare, {\it {Two-dimensional S-matrices from
  unitarity cuts}},  {\em JHEP} {\bf 1307} (2013) 088,
  [\href{http://xxx.lanl.gov/abs/1304.1798}{{\tt arXiv:1304.1798}}].

\bibitem{Bianchi:2014rfa}
L.~Bianchi and B.~Hoare, {\it {$AdS\_3 \times S^3 \times M^4$ string S-matrices
  from unitarity cuts}},  {\em JHEP} {\bf 08} (2014) 097,
  [\href{http://xxx.lanl.gov/abs/1405.7947}{{\tt arXiv:1405.7947}}].

\bibitem{Engelund:2014pla}
O.~T. Engelund and R.~Roiban, {\it {On the asymptotic states and the quantum S
  matrix of the $\eta$-deformed AdS$_{5} \times$ S$^{5}$ superstring}},  {\em
  JHEP} {\bf 03} (2015) 168, [\href{http://xxx.lanl.gov/abs/1412.5256}{{\tt
  arXiv:1412.5256}}].

\bibitem{Hoare:2014kma}
B.~Hoare, A.~Pittelli, and A.~Torrielli, {\it {Integrable S-matrices, massive
  and massless modes and the AdS$_{2}$ * S${2}$ superstring}},  {\em JHEP} {\bf
  11} (2014) 051, [\href{http://xxx.lanl.gov/abs/1407.0303}{{\tt
  arXiv:1407.0303}}].

\end{thebibliography}\endgroup

\end{document}